\documentclass[aip,preprint,groupedaddress,floatfix]{revtex4-2}

\usepackage{graphicx}
\usepackage{amsmath}
\usepackage{amssymb}
\usepackage{bm}
\usepackage{booktabs}
\usepackage{xcolor}
\usepackage{array}[=2016-10-06]
\usepackage{siunitx}
\usepackage[T1]{fontenc}
\usepackage{mathptmx}
\usepackage{hyperref}
\usepackage{capt-of}
\usepackage{listings}
\newsavebox{\budgettab}

\begin{document}

\title{Co-design of Silicon Microring Modulator beyond 200~Gb/s per Lane: Device Physics,
       Operating Point, and Compact Models
       for Scale-Up and Scale-Out Optical I/O}

\author{Zhihong Huang}
\email{zhihong.huang@hpe.com}
\affiliation{Hewlett Packard Labs, Hewlett Packard Enterprise, Milpitas, CA 95035, USA}

\author{Yuan Yuan}
\affiliation{Hewlett Packard Labs, Hewlett Packard Enterprise, Milpitas, CA 95035, USA}

\author{Yiwei Peng}
\affiliation{Hewlett Packard Labs, Hewlett Packard Enterprise, Milpitas, CA 95035, USA}

\author{Samuel Palermo}
\affiliation{Department of Electrical and Computer Engineering,
             Texas A\&M University, College Station, TX 77843, USA}

\author{Marco Fiorentino}
\affiliation{Hewlett Packard Labs, Hewlett Packard Enterprise, Milpitas, CA 95035, USA}

\author{Raymond G. Beausoleil}
\affiliation{Hewlett Packard Labs, Hewlett Packard Enterprise, Milpitas, CA 95035, USA}

\date[]{}

\begin{abstract}



As artificial intelligence clusters scale to millions of accelerators, 
computing throughput is increasingly limited by the bandwidth density 
and energy efficiency of the optical interconnect fabric. 
Depletion-mode silicon microring modulators provide an essential path forward, combining femtojoule-per-bit junction switching energy with dense wavelength-division multiplexing. At these baud rates, device physics interacts directly with driver electronics, receiver noise, and link equalization. This review establishes a device-to-link pathway from 200~Gb/s PAM4 transmitters to the emerging 400G-per-lane frontier, showing that the optical operating point is an end-to-end link trade-off rather than an isolated modulator property. To enable unified electronic--photonic co-design, we evaluate how optical self-heating, free-carrier dispersion, and ambient temperature drift shift the resonance during modulation, setting quantitative requirements on cold-resonance placement and thermal tuning reserves. We deliver a calibrated physical and Verilog-A modeling framework that captures these coupled dynamics in circuit simulation, equipping system architects and transceiver designers to optimize 200G optical engines with predictable margins while establishing a foundation for 400G-per-lane architectures.

\end{abstract}

\maketitle

\section{Introduction}\label{sec:intro}

The scaling of artificial intelligence systems is constrained by the bandwidth and
energy cost of moving data among accelerators. Training a frontier model requires
tens of thousands of accelerators to operate as a single machine, but their
computational throughput has grown faster than the bandwidth available for
communication. Consequently, accelerator utilization can be limited by data
movement among processors, memory and racks more than by
computation~\cite{gholami2024memory,metz2024energy}.

Meeting this communication demand with electrical interconnects becomes more
difficult as per-lane rates rise from 100 toward 200~Gb/s and beyond. Signals
propagating through copper package and board channels undergo frequency-dependent
attenuation, crosstalk and reflections, requiring equalization to compensate for
channel loss and intersymbol interference (ISI). Increasing the number of lanes
cannot remove these constraints indefinitely, because switch front-panel area and
accelerator input/output (I/O) perimeter also limit the available connections.

Optical interconnects offer a route beyond these electrical reach constraints,
with low fiber propagation loss over intra- and inter-rack distances. Their
integration into computing systems has been supported by two decades of
silicon-photonics development, through which modulators, photodetectors and passive
circuits have reached volume manufacturing in CMOS-compatible
processes~\cite{siew_jlt2021,shekhar_nc2024,margalit_apl2021}. To shorten the
remaining electrical path, optical engines are moving from faceplate pluggables
toward near-package and co-packaged optics, millimeters from the switch or accelerator
die~\cite{minkenberg_iet2021,tan_foe2023,nagarajan_jstqe2022,abrams_jlt2020}.
Coupling and component losses nevertheless remain part of the optical power budget.

Replacing the transmission medium does not, however, eliminate the bandwidth and
energy constraints within the transceiver. At 200~Gb/s per lane, the modulator,
photodetector and associated circuits can limit the link response, so signal
processing remains necessary to achieve the target error rate. Multi-tap
equalization and forward-error correction (FEC) in the DSP consume electrical
power and add latency. The energy advantage of optical transmission therefore
depends on how the device response and electronic signal processing are designed
together, not on propagation loss alone.


Over the last decade, the Large-Scale Integrated Photonics group at Hewlett Packard
Labs has developed a family of silicon photonics components for such
links~\cite{huang_optica2016,liang_jstqe2022,yuan_nc2024,peng_np2024,kurczveil_ptl2018,zhang_optica2019}.
On the optical source side, heterogeneous hybrid-bonded III--V-on-silicon microring lasers
and quantum-dot frequency comb lasers provide energy-efficient, multi-wavelength
optical carriers across dense channel grids~\cite{liang_jstqe2022,zhang_optica2019,kurczveil_ptl2018}.
High-speed depletion-mode silicon
microring modulators use resonant enhancement to obtain low-voltage modulation,
supporting 200~Gb/s PAM4 operation with 49~GHz of electro-optic bandwidth
and femtojoule-scale junction switching energy per bit~\cite{yuan_nc2024}. Their performance nevertheless
depends on the interplay between junction efficiency, cavity lifetime and laser
detuning. This review focuses on those device-level trade-offs and the operating
conditions needed to obtain useful modulation. On the receiver side, silicon/silicon-germanium
APDs provide internal avalanche multiplication gain, which can improve sensitivity and reduce
the required optical power~\cite{huang_ofc2020,huang_ecoc2025_invited}; the avalanche receiver and
link-level co-design are treated in a companion review.

This review is organized by first placing the transmitter within the landscape of AI
interconnects, where scale-up and scale-out fabrics impose different requirements
(Sec.~\ref{sec:landscape}). Junction and cavity physics are then used to relate
coupling and laser detuning to modulation above 200~Gb/s per lane, distinguishing
the operating points with optimal transmission power (Sec.~\ref{sec:transceiver}).
Because optical absorption also shifts the resonance, these operating points must
be considered together with nonlinear optical effects and heater tuning
(Secs.~\ref{sec:selfheat} and \ref{sec:mrmnl}). The electrical, optical and thermal
equations provide a basis for translation into Verilog-A for system level simulation. Numerically
calculated PAM4 eyes show how device response and operating point affect the
transmitted waveform (Sec.~\ref{sec:va}). 
\section{The AI optical-interconnect landscape}\label{sec:landscape}

\begin{figure*}[t]
  \centering
  \includegraphics[width=\textwidth]{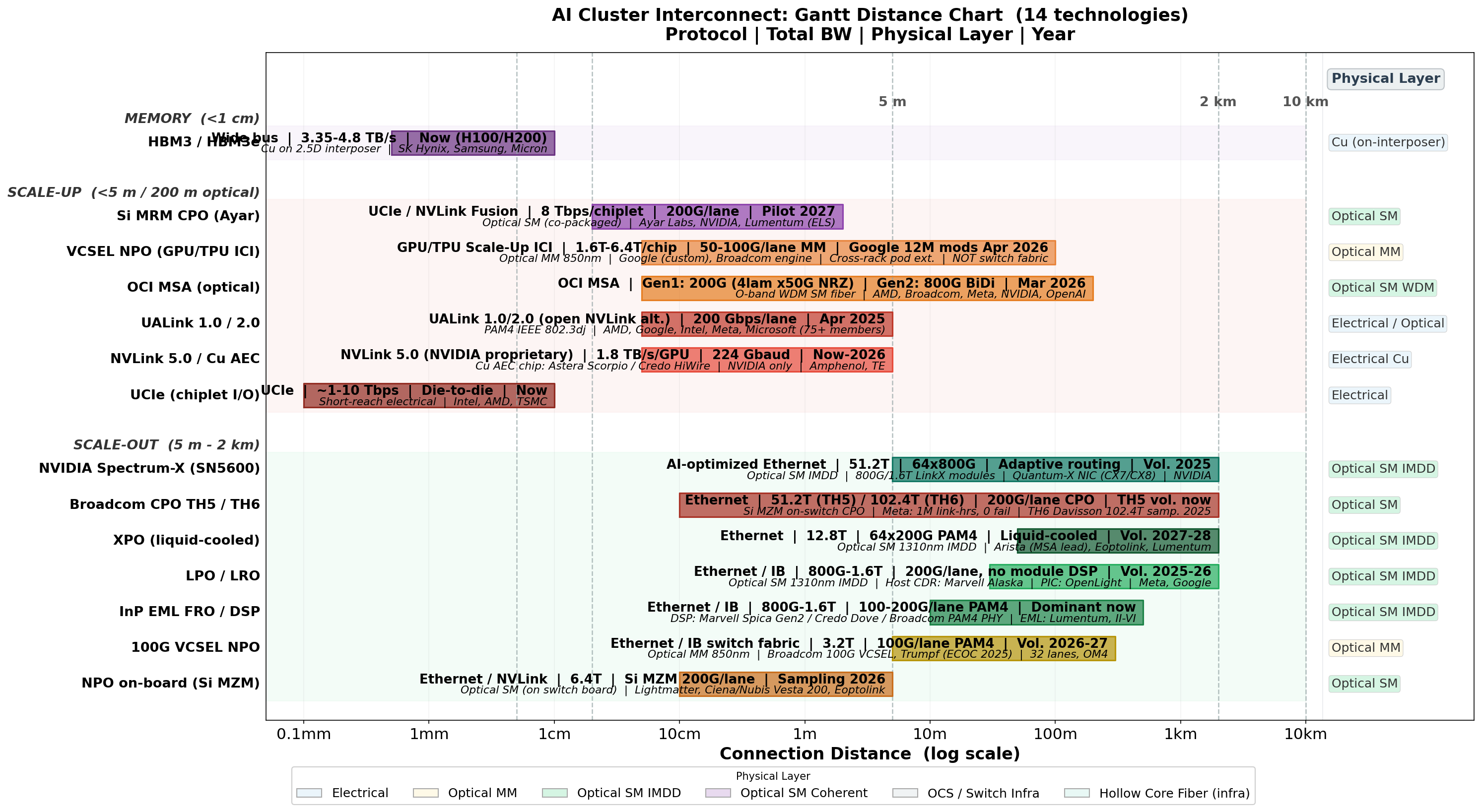}
  \caption{AI interconnect technologies organized by connection distance and physical layer.}
  \label{fig:landscape}
\end{figure*}

The interconnect architecture of artificial intelligence computing clusters is driven by
fundamental memory and bandwidth scaling limits. Frontier neural network models, spanning hundreds of
billions to trillions of parameters, exceed the memory capacity and compute throughput
of any single accelerator chip. Distributing training across thousands of processors splits
communication into distinct physical domains categorized by reach and loss constraints (Fig.~\ref{fig:landscape}).
Scale-up links span millimeters on-package to several meters within a server chassis or rack,
where tensor and pipeline parallelism demand continuous, low-latency all-reduce exchanges constrained
by shoreline density and energy dissipation.
Scale-out fabrics extend from 1~m to 2~km (and up to 10~km in campus networks), connecting
accelerators and switch tiers across racks and rows under strict optical loss budgets.
Beyond the cluster, scale-across networks span tens to thousands of kilometers between
distant datacenters using coherent optical transport.
Whether an optical link serves scale-up or scale-out determines the optimum modulator junction design,
drive swing, and optical operating point.

\subsection{Scale-up fabrics and shoreline constraints}
Scale-up fabrics link accelerators into a single, tightly synchronized logical computer,
typically within a server chassis or rack. Current accelerators exchange tensor parameters
over proprietary fabrics such as NVLink or open standards such as UALink. Emerging die-to-die
interfaces such as UCIe and memory protocols such as CXL over PCIe address complementary
on-package and host-attached communication.
At a lane rate of 200~Gb/s per unidirectional channel, an aggregate bidirectional bandwidth
of 1.8~TB/s requires 72 physical channels per package, or 36 in each direction for a symmetric
interface; a 3.6~TB/s target requires 144 channels in total.
During distributed neural network training, collective communication routines such as
All-Reduce, All-Gather, and Reduce-Scatter require all participating processors to exchange
intermediate gradients before subsequent forward and backward computation can proceed.
Because every processor must complete its exchange before the collective barrier clears,
tail latency and link-to-link skew directly stall compute pipelines, making high-latency,
power-intensive digital signal processors (DSPs) unacceptable in scale-up networks.

Electrical channels traversing packages and printed circuit boards cannot support these
data rates across meter-scale distances. At symbol rates exceeding 100~GBaud, copper traces
suffer severe dielectric attenuation, crosstalk, and impedance discontinuities, restricting
high-speed electrical reach to tens of centimeters. Thermal dissipation and package physical
geometry impose further restrictions. High-Bandwidth Memory (HBM) stacks occupy nearly the
entire perimeter of the accelerator package, leaving millimeter-scale shoreline windows for
external input/output (I/O).
Escaping terabytes per second through this congested edge requires ultra-high bandwidth
density in terabits per second per millimeter of package shoreline.

Depletion-mode silicon microring modulators provide an ideal physical match for scale-up
fabrics. Their compact footprint and small junction
capacitance enable multi-channel dense wavelength-division multiplexing (DWDM) across
8 to 16 wavelengths on a single bus waveguide and single fiber, without requiring bulky
arrayed waveguide gratings or external optical multiplexers.
Furthermore, the small junction area yields switching energies below 100~fJ/bit, enabling
direct driving from low-voltage CMOS electronics without power-hungry retimers.
However thermal stability remains a chanllenge for microrings. Compute accelerators
dissipate hundreds of watts, generating severe localized thermal flux and dynamic junction
temperature excursions. Under the OIF 3.2~Tb/s co-packaged optics implementation
agreement~\cite{oif_cpo_module2023}, optical engines without integrated lasers operate across
a case temperature range of 15 to 85~\textcelsius.
With silicon exhibiting an 80~pm/K thermo-optic resonance shift~\cite{xu_oe2019}, uncompensated
temperature drift detunes the optical cavity from the laser wavelength, degrading modulation
depth and closing the optical eye.
Consequently, scale-up transmitters require integrated micro-heaters and active closed-loop
thermal stabilization~\cite{padmaraju_nanoph2014}. 


\subsection{Scale-out fabrics and optical link budgets}
Scale-out fabrics connect switch application-specific integrated circuits (ASICs) and server nodes across racks,
rows, and datacenter clusters, spanning reaches from 1~m to 2~km, and up to 10~km for campus-wide links.
Pluggable optical transceivers, near-package optics (NPO), and co-packaged optics (CPO) differ
in the physical separation between the optical engine and the switch ASIC. Moving the optics
closer to the switch silicon shortens the host electrical channel and reduces electrical energy
dissipation, while pluggable transceivers offer modularity and straightforward field
servicing~\cite{minkenberg_iet2021,tan_foe2023,nagarajan_jstqe2022}.

In scale-out interfaces, the high-speed electrical signal travels across the host printed circuit
board (PCB) to the optical transceiver or co-packaged engine, where it converts into an optical signal.
The optical waveform then propagates through optical connectors and single-mode fibers across racks
and switch tiers to reach destination servers.
Unlike scale-up links where package shoreline density dominates, scale-out links are governed by the
optical channel loss budget. Optical connector insertions and fiber attenuation across 100~m to 2~km
paths introduce 4 to 6~dB of optical channel loss in standard 200GBASE-DR4 and FR4 specifications.
To close this link budget under interoperability standards such as IEEE P802.3dj
for 200~Gb/s per lane~\cite{welch_ieee2024} and OIF CEI-224G~\cite{oif_cei224g}, the optical
transmitter must deliver high launch optical modulation amplitude (OMA) while satisfying strict
transmitter and dispersion eye closure quaternary (TDECQ) limits.

In this loss-dominated regime, competing optical modulator technologies present distinct
performance trade-offs. Indium phosphide (InP) electro-absorption modulators (EAMs) have
demonstrated 155~GBaud PAM4 at 310~Gb/s and 400~Gb/s using PAM6~\cite{uchiyama_ofc2024}, and
heterogeneous InP modulators on silicon have achieved 170~GBaud PAM4 over 500~m of optical
fiber~\cite{ostrovskis_ofc2024}. Thin-film lithium niobate (TFLN) modulators deliver electro-optic
bandwidths beyond 100~GHz with low drive voltages and high linearity~\cite{wang_nature2018,xu_nc2020}.
Plasmonic modulators provide wide analog bandwidth and femtofarad capacitance at the cost of
high insertion loss~\cite{burla_app2019,heni_nc2019}. Silicon Mach--Zehnder and microring modulators must likewise meet the bandwidth and drive-voltage requirements for operation at these rates.

When deployed in scale-out fabrics, 
the microring transmitter can no longer be operated solely for low drive swing.
Instead, the operating point must shift toward
the maximum dynamic OMA regime, where optical cavity peaking deliberately
compensates for high-frequency electrical roll-off. Although this operating point introduces
transient overshoot and higher raw eye distortion, receive-side feed-forward equalization
(FFE) or digital signal processing recovers the eye margin~\cite{berikaa_oft2022}.

Beyond the cluster switch fabric, scale-across and long-haul links connect distributed
datacenters across metropolitan and continental distances. These networks use coherent optical
transceivers encoding dual-polarization amplitude and phase under OIF 800ZR and 1600ZR
implementation agreements~\cite{oif_800zr,oif_1600zr}.
Because coherent systems rely on external local oscillators, complex IQ modulators, and
full-retiming DSPs, their receiver architecture differs fundamentally from the direct-detection
intensity-modulated links addressed here.
The divergent constraints across these interconnect regimes motivate the integrated
transceiver architecture and analytical modeling framework developed in Section~\ref{sec:transceiver}.
\section{Analytical modeling of silicon microring modulators}\label{sec:transceiver}

The divergent constraints of scale-up and scale-out demonstrate that an optical modulator
cannot be evaluated or optimized as an isolated device. In scale-up fabrics, high shoreline
bandwidth density, low driver dissipation, and self-restoring thermal stability govern the
transceiver, favoring compact, low-capacitance resonators. In scale-out fabrics, launch optical
modulation amplitude and receiver sensitivity dominate, requiring transmitters to close multi-decibel
optical loss budgets across hundreds of meters to kilometers of fiber. Translating these
conflicting network demands into quantitative device specifications requires analyzing the
modulator, channel, and receiver as an integrated link.

\begin{figure*}[t]
  \centering
  \includegraphics[width=\textwidth]{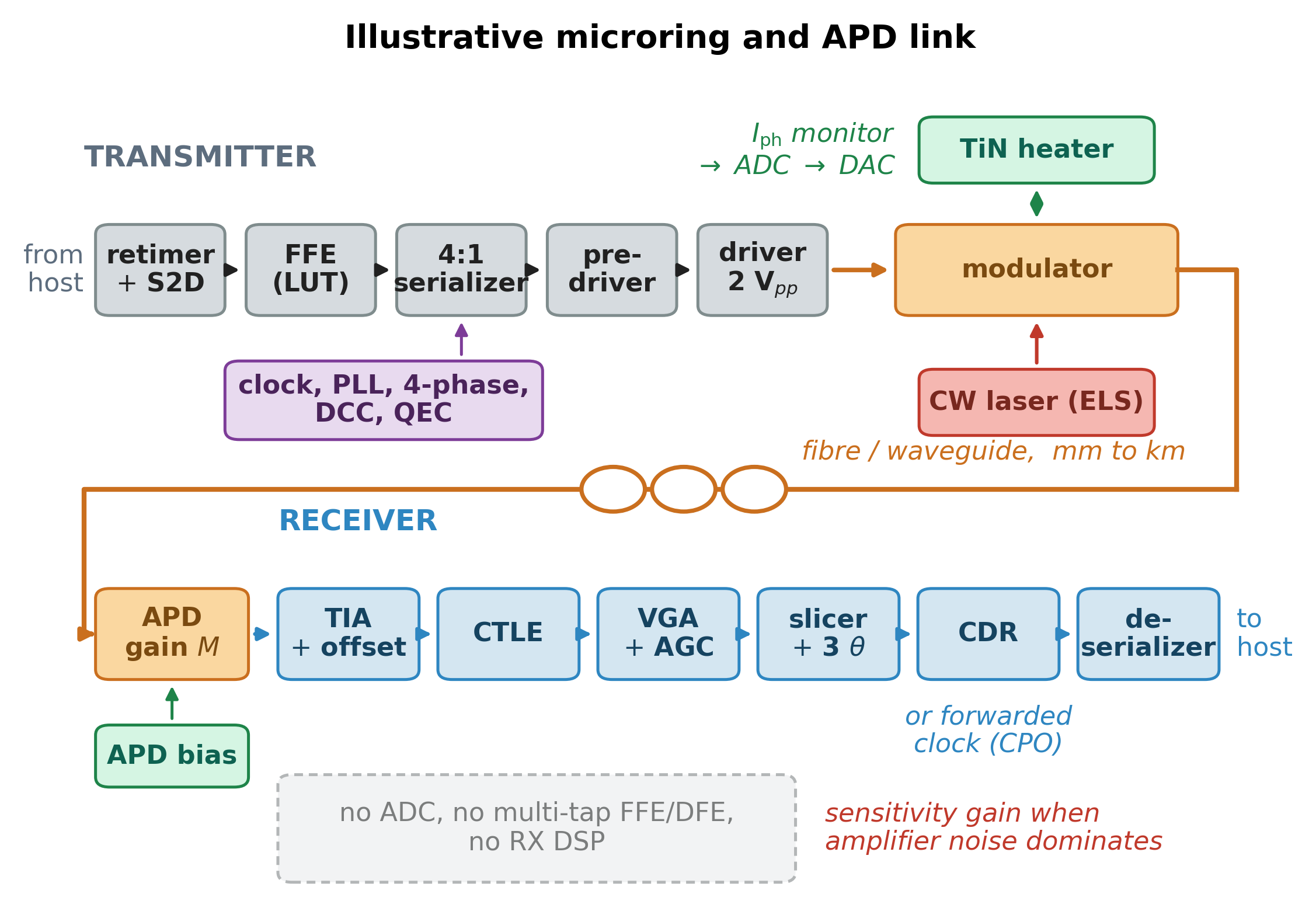}
  \caption{Illustrative MRM--APD link with analog receiver equalization
  and slicing. Thermal-loop converters are outside the data path.}
  \label{fig:link}
\end{figure*}

Figure~\ref{fig:link} shows an end-to-end optical link combining a depletion-mode silicon
microring modulator with an avalanche photodiode (APD) receiver. The depletion-mode microring
modulator converts a voltage-induced resonance shift into an intensity change when biased on a
resonance flank. Resonant enhancement enables high-speed modulation within a sub-15~$\mu\mathrm{m}$
cavity, reducing active junction length and lowering capacitive switching energy to femtojoule-per-bit
levels. This switching-energy advantage must be evaluated against the optical modulation
amplitude (OMA) and bandwidth delivered to the channel, accounting for junction parasitics,
driver swing, and optical insertion loss.

On the receive side, the APD provides internal avalanche multiplication gain $M$ before the
transimpedance amplifier, suppressing the input-referred thermal noise of the electronic front
end. When electronic amplifier noise dominates, avalanche multiplication improves receiver
sensitivity, directly relaxing the transmitter launch OMA required to achieve the target bit-error
rate across the link. Conversely, in low-loss scale-up links where optical power is preserved,
optical engines can prioritize low driver swing and minimal equalization latency. The analytical
framework developed in the following sections resolves these interrelated device, cavity, and
circuit dynamics, beginning with junction design and active carrier-mode overlap.

\subsection{High-speed silicon microring modulators}\label{sec:devices}

Si depletion-mode microring modulators have demonstrated
128~Gb/s PAM4 with integrated thermo-optic
tuning~\cite{sun_jlt2019} and 240~Gb/s on lateral~\cite{zhang_prj2022}
and L-shaped junction platforms~\cite{sakib_ofc2022}.
More recent work reports 400~Gb/s PAM6 eyes with DSP and acquisition
averaging~\cite{hu_arxiv2025}. 

Table~\ref{tab:ringdesigns} compares four
high-speed Si microring modulators, including
our 5$\times$200~Gb/s PAM4 Z-shaped microring modulator~\cite{yuan_nc2024},
an eight-channel PAM4 transmitter using 256G lateral rings~\cite{xue_ofc2025},
a 16$\times$128~Gb/s device~\cite{bu_ofc2026} and a lateral two-segment
device~\cite{xue_jlt2025}. The devices differ in junction overlap, operating point,
and the balance between per-lane rate and channel count.

\begin{figure*}[t]
  \centering
  \includegraphics[width=\textwidth]{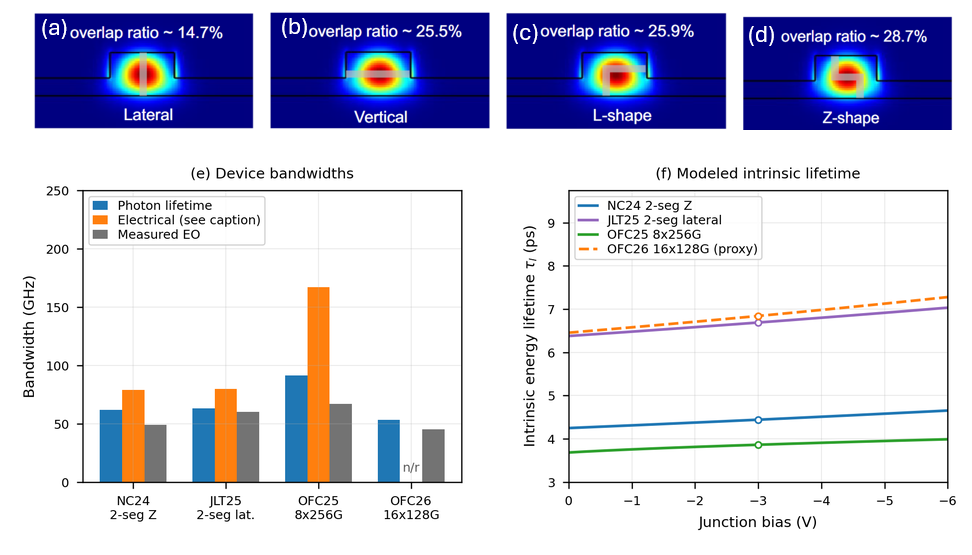}
  \caption{Junction comparison. \textbf{(a)--(d)}~Simulated modes and
  depletion overlaps. \textbf{(e)}~Photon-lifetime, electrical and measured
  EO bandwidths from Table~\ref{tab:ringdesigns}, taking the shorter-segment
  electrical value for each two-segment device, where n/r denotes not
  reported. \textbf{(f)}~Modeled intrinsic energy
  lifetime versus bias. Circles mark assumed $-3$~V references; the dashed
  128~Gb/s curve uses a doping proxy.}
  \label{fig:junction}
\end{figure*}

The junction cross-section sets the overlap $\Gamma$ between the
optical mode and the depletion region. At a common depletion width of 40~nm,
the calculated overlaps are $\sim$14.7\% for a lateral junction,
$\sim$25.5\% for a vertical junction, $\sim$25.9\% for an L-shaped junction
and $\sim$28.7\% for a Z-shaped junction
[Fig.~\ref{fig:junction}(a)--(d)]~\cite{yuan_nc2024}. A larger overlap improves the
modulation efficiency at a given junction capacitance. Compared with a lateral
junction of $V_\pi L \sim 1.00$~V$\cdot$cm~\cite{yuan_acsp2022}, the Z-shaped junction
reaches $\sim$0.67~V$\cdot$cm, corresponding to higher phase-modulation
efficiency, and doubles the data rate at
$\sim$36\% lower junction-switching energy per bit~\cite{yuan_nc2024}. The three devices in
Table~\ref{tab:ringdesigns} with reported $V_\pi L$ values span a narrow range
of 0.63--0.69~V$\cdot$cm despite their different junction geometries.
Reduced optical overlap can be compensated by higher junction doping, at the
expense of increased junction capacitance per unit length.
Increasing reverse bias reduces carrier absorption and lengthens the
intrinsic energy lifetime $\tau_l$ [Fig.~\ref{fig:junction}(f)].

Junction doping sets the depletion width and hence both the tuning efficiency
and $C_j$. Higher doping narrows the depletion region and increases the
capacitance, while the usable reverse-bias swing remains constrained by electric
field, leakage and breakdown. The doping profile must therefore balance tuning
efficiency, capacitance and optical loss at the intended drive conditions.
The Z-shaped ring's upper junction region has p-type and n-type concentrations of
$3\times10^{18}$~cm$^{-3}$ each~\cite{yuan_nc2024}. The 256G lateral ring
in the eight-channel transmitter uses an n-type concentration of
$10^{19}$~cm$^{-3}$ and a p-type concentration of
$5.3\times10^{18}$~cm$^{-3}$~\cite{xue_ofc2025}.
The reconstructed series resistance is $\sim$16~$\Omega$ for the
256G lateral ring in the eight-channel transmitter, compared with a fitted $\sim$386~$\Omega$
for the shorter segment of the lateral two-segment ring~\cite{xue_jlt2025},
as specified in Table~\ref{tab:ringdesigns}.

Table~\ref{tab:ringdesigns} groups the devices by junction segmentation.
Single-segment microring modulators drive a continuous junction
around the ring. Direct PAM4 modulation requires a four-level electrical
waveform, commonly generated by a DAC and driver, with increasing power and
equalization demands at higher rates~\cite{xue_ofc2025}. Two-segment designs
instead generate four optical levels from two NRZ drives for the least
significant bit (LSB) and most significant bit (MSB), using nominal
modulation weights of one and two~\cite{yuan_acsp2022,yuan_nc2024}.
Segmentation also reduces the load on each driver. The Z-shaped segments
have capacitances of $\sim$6.6 and $\sim$13.2~fF, against $\sim$19.8~fF for an equivalent
one-segment junction. Including the source and fitted parasitic network, the
corresponding electrical bandwidths increase from $\sim$54~GHz for the
one-segment case to $\sim$79 and $\sim$65~GHz for the shorter and longer
segments, respectively~\cite{yuan_nc2024}. The bandwidth improvement follows
from the smaller segment loads within the fitted network.

The Z-shaped ring combines
enhanced junction overlap with two small capacitive loads, demonstrating
100~GBaud PAM4 with two 1.6~V binary drives. The 256G lateral
ring reaches 128~GBaud with a 1.8~V PAM4 drive
(Table~\ref{tab:ringdesigns}). These architectures connect junction
efficiency, cavity coupling and driver loading to the requirements for
200G-class operation.

\begin{table}[t]
\centering
\caption{Microring device comparison. Paired values denote shorter/longer
segments; swings are per segment and bias points denote laser detuning.}
\label{tab:ringdesigns}
\small
\setlength{\tabcolsep}{4.5pt}
\renewcommand{\arraystretch}{1.15}
\begin{tabular}{lcccc}
\toprule
 & \multicolumn{2}{c}{\textbf{Two-segment}} & \multicolumn{2}{c}{\textbf{Single-segment lateral}} \\
\cmidrule(lr){2-3}\cmidrule(lr){4-5}
\textbf{Parameter} & \textbf{Z-shaped}~\cite{yuan_nc2024}
                   & \textbf{Lateral}~\cite{xue_jlt2025}
                   & \textbf{8\,$\times$\,256~Gb/s}~\cite{xue_ofc2025}
                   & \textbf{16\,$\times$\,128~Gb/s}~\cite{bu_ofc2026} \\
\midrule
Ring radius (\textmu m)       & 12            & 6            & 6            & 6            \\
Loaded $Q_L$                    & 3700          & 3700         & 2500         & 4300         \\
DC extinction ratio (dB)      & 16            & ---          & $>$20        & $>$18        \\
Coupling condition            & ---           & over (12\% target) & near-critical & near-critical \\
Tuning efficiency (pm/V)      & 27            & 7 / 15       & 33           & 27           \\
$V_\pi L$ (V$\cdot$cm)        & 0.67          & 0.63         & 0.69         & ---          \\
$C_j$ (fF)                    & 6.6 / 13.2    & 6.0 / 10.0   & $\sim$20$^*$ & ---          \\
$C_\mathrm{pad}$ (fF)         & 31.6 / 33.1   & 24.2 / 24.6  & $\sim$36$^*$ & ---          \\
$R_s$ ($\Omega$)              & 68.1 / 35.4   & 385.9 / 246.7 & $\sim$16$^*$ & ---          \\
Electrical bandwidth (GHz)    & 79 / 65       & 80 / 75      & 167$^*$      & ---          \\
$f_\mathrm{ph}$ (GHz)         & 62            & 63           & 92           & 53           \\
$f_\mathrm{EO}$ (GHz)         & 49 / 48       & 60           & 67           & 45           \\
Bias point                    & max OMA       & near max GBW & near max GBW & near max GBW \\
Drive swing (V$_\mathrm{pp}$) & 1.6           & $\sim$2      & 1.8          & 1.6          \\
Symbol rate (GBaud)           & 100           & 64           & 128          & 64           \\
TDECQ (dB)                    & ---           & ---          & 2.6--2.9     & 2.7--2.9     \\
\bottomrule
\end{tabular}
\par\smallskip
{\footnotesize\raggedright
$^*$Calculated from this paper.
\par}
\end{table}


\subsection{Depletion-mode silicon microring modulators}\label{sec:mrm}


\subsubsection{Coupled-mode theory}
A microring is resonant when its round-trip phase is an
integer number of cycles~\cite{bogaerts_lpr2012,van_book2016}. Interference
between the bus and circulating fields converts resonance tuning into
intensity modulation. The coupled-mode description relates resonator--waveguide
coupling~\cite{yariv_el2000} to RF-to-optical conversion~\cite{ehrlichman_oe2018},
with the cavity response~\cite{bao_jlt2024}
\begin{equation}
  \frac{da(t)}{dt} = \Bigl[\,j\,\omega_r - \frac{1}{2\tau_p}\,\Bigr] a(t)
                  + \frac{1}{\sqrt{\tau_e}}\; E_\mathrm{in}(t),
  \qquad
  E_\mathrm{out}(t) = E_\mathrm{in}(t) - \frac{1}{\sqrt{\tau_e}}\; a(t),
  \label{eq:cmt_mrm}
\end{equation}
where $a(t)$ is the cavity amplitude and $|a(t)|^2$ the stored energy.
The bus fields $E_\mathrm{in}(t)$ and $E_\mathrm{out}(t)$ are normalized
so that $|E(t)|^2$ is optical power. The resonance angular frequency is
$\omega_r=2\pi c/\lambda_\mathrm{res}$, where $\lambda_\mathrm{res}$ is
the resonance wavelength.
The energy lifetimes $\tau_e$ and $\tau_l$ describe coupling and
intrinsic loss, with combined photon lifetime $\tau_p$. After removing
the input at $t=0$, fixed $\omega_r$ and $\tau_p$ give
\begin{equation}
 \begin{aligned}
  a(t) &= a(0)e^{j\omega_r t}e^{-t/(2\tau_p)}, 
  \qquad
  |a(t)|^2 &= |a(0)|^2e^{-t/\tau_p}.
 \end{aligned}
  \label{eq:free_decay}
\end{equation}
The stored energy falls to $1/e$ of its initial value after $\tau_p$,
whereas the field magnitude requires $2\tau_p$. Longer storage narrows
the resonance, relating these lifetimes to the quality factors,
\begin{equation}
  Q_L \equiv \frac{\lambda_\mathrm{res}}{\Delta\lambda_\mathrm{FWHM}} = \omega_r\tau_p
      = \frac{2\pi c\tau_p}{\lambda_\mathrm{res}},
  \qquad
  Q_i = \omega_r\tau_l, 
  \qquad
  Q_c = \omega_r\tau_e
  \label{eq:static}
\end{equation}

where $Q_L$, $Q_i$ and $Q_c$ are the loaded, intrinsic and coupling
quality factors, and $\Delta\lambda_\mathrm{FWHM}$ is the full width at half the notch
depth in linear power. The two energy-decay rates add, giving
\begin{equation}
 \begin{aligned}
  f_\mathrm{ph}
  &= \frac{c}{\lambda_\mathrm{res}Q_L}
  = \frac{1}{2\pi \tau_p},
  \qquad
  \frac{1}{\tau_p}
   = \frac{1}{\tau_e}+\frac{1}{\tau_l}.
 \end{aligned}
  \label{eq:loadeddecay}
\end{equation}
where $f_\mathrm{ph}$ is the linewidth-derived photon-lifetime frequency
used in the approximate bandwidth budget.
A 67~GHz EO target requires $\tau_p<2.4$~ps within the
bandwidth budget of Eq.~(\ref{eq:bw}), with additional allowance for
junction charging. At 1310~nm, the Z-shaped ring's loaded quality factor
$Q_L=3700$ gives 2.6~ps and 62~GHz, compared with 1.7~ps and
92~GHz for the 256G lateral ring at $Q_L=2500$. Their measured EO bandwidths
are 49/48~GHz for the Z-shaped segments and 67~GHz for the
256G lateral ring (Table~\ref{tab:ringdesigns}). Enhanced junction overlap provides
efficient tuning, while stronger bus coupling shortens the photon lifetime.
The intrinsic lifetime $\tau_l$ is determined by free-carrier absorption,
scattering and bending radiation. Reverse-bias depletion reduces
absorption and lengthens this lifetime, while self-heating changes both
loss and resonance position (Sec.~\ref{sec:selfheat}). The external
lifetime $\tau_e$ depends on the energy transferred to the bus per round trip.

At resonance, the extent of destructive interference
is determined by the balance between external coupling and intrinsic loss,
\begin{equation}
 \begin{aligned}
  \rho &\equiv \frac{\tau_l}{\tau_e} = \frac{Q_i}{Q_c}, 
  \qquad
  D &= 1 - \left(\frac{1-\rho}{1+\rho}\right)^{2}, 
  \qquad
  \mathrm{ER}_\mathrm{notch} &= -10\log_{10}(1-D).
 \end{aligned}
  \label{eq:notchdepth}
\end{equation}
where $\rho$ is the ratio of external to intrinsic decay rates, $D$ the
fractional on-resonance power extinction, and $\mathrm{ER}_\mathrm{notch}$
the static notch ER. Critical coupling gives complete cancellation in
this ideal model. The Z-shaped ring has a 16~dB notch ER, while the
256G lateral ring exceeds 20~dB, requiring
$D\geq99\%$. On the slightly
overcoupled branch where the coupling $Q_c$ is less than the intrinsic $Q_i$, 
stronger coupling broadens the
resonance at the expense of notch depth. At $\rho=1.2$ and $Q_L=2500$,
the intrinsic and coupling quality factors $Q_i$ and $Q_c$ are approximately
5500 and 4583, respectively. The coupler power transfer is related to the external lifetime by

\begin{equation}
\kappa^{2} \simeq \frac{n_gL_\mathrm{rt}}{c\tau_e},
\qquad
\kappa^{2} + |t_c|^{2} = 1,
\label{eq:kappa}
\end{equation}
where $\kappa^2$ is the coupler power-transfer fraction, $n_g$ the group index
and $L_\mathrm{rt}$ the round-trip length. The lifetime approximation
assumes weak coupling, and power conservation assumes a lossless coupler
with field self-coupling coefficient $t_c$.

The photon lifetime sets the wavelength linewidth,
\begin{equation}
  \Delta\lambda_\mathrm{FWHM} \simeq \frac{\lambda_\mathrm{res}^{2}}{2\pi c\,\tau_p}.
\end{equation}
The round-trip group delay, independent of the decay time, sets the spacing
between adjacent resonances,
\begin{equation}
  \mathrm{FSR} \simeq \frac{\lambda_\mathrm{res}^{2}}{n_g L_\mathrm{rt}},
  \label{eq:fsr}
\end{equation}
where FSR is the local wavelength spacing. For the 256G lateral ring,
$n_g=4.0$ and $L_\mathrm{rt}=38.7$~$\mu$m give an 11.1~nm FSR at 1310~nm.

At a given group index, the target FSR fixes the round-trip length. A circular
ring and a racetrack with two equal straight sections then satisfy
\begin{equation}
  R_\mathrm{ring} \simeq \frac{\lambda_\mathrm{res}^{2}}{2\pi n_g\,\mathrm{FSR}}
  \quad\text{(circular ring)},
  \qquad
  L_\mathrm{rt}=2\pi R_\mathrm{ring}+2L_s
  \quad\text{(racetrack)},
  \label{eq:ringgeometry}
\end{equation}
where $R_\mathrm{ring}$ is the bend radius and $L_s$ is the length of each
straight section. At fixed FSR and group index, a longer straight
interaction requires smaller bends, linking coupling length to bending loss.

For a uniform, lossless, phase-matched straight coupler of length $L_c$,
the even and odd supermodes give
\begin{equation}
  \kappa^2=\sin^2\!\left[\frac{(\beta_+-\beta_-)L_c}{2}\right],
  \label{eq:couplinglength}
\end{equation}
where $\beta_+$ and $\beta_-$ are the propagation constants of the even
and odd supermodes. A smaller gap generally increases their splitting,
and weak-transfer coupling grows quadratically with interaction length.
Approach bends and phase mismatch must be included in the fabricated design.

For monochromatic input of wavelength $\lambda$ and fixed cavity parameters,
the cavity-only through-port field transmission is
\begin{equation}
  t(\lambda) = \frac{E_\mathrm{out}(t)}{E_\mathrm{in}(t)}
    = 1 - \frac{1/\tau_e}
        {1/(2\tau_p) - j\,2\pi c(1/\lambda_\mathrm{res}-1/\lambda)},
  \label{eq:tfield}
\end{equation}
where $\lambda$ is the laser wavelength and $t(\lambda)$ approaches unity
far from resonance. The fields are defined at the bus ports in
Eq.~(\ref{eq:cmt_mrm}), with resonance-minus-laser detuning
$\Delta\omega=\omega_r-\omega_\mathrm{in}
=2\pi c(1/\lambda_\mathrm{res}-1/\lambda)$.

Including nonresonant loss between the chosen input and output reference
planes gives the absolute power transmission $T(\lambda)$. With detuning
normalized to the resonance half-linewidth,
\begin{equation}
  x \;\equiv\; 2\Delta\omega\,\tau_p
    \;\simeq\; \frac{2\,(\lambda-\lambda_\mathrm{res})}{\Delta\lambda_\mathrm{FWHM}},
  \qquad
  T(\lambda) = T_\mathrm{off}|t(\lambda)|^2 = T_\mathrm{off}\left[1 - \frac{D}{1+x^{2}}\right],
  \label{eq:xnorm}
\end{equation}
where $x$ is the dimensionless laser detuning and
$T_\mathrm{off}=|t_\mathrm{off-res}|^2$ is the off-resonance power transmission.
The off-resonance field ratio $t_\mathrm{off-res}$ includes waveguide and
taper losses and is treated as constant across the resonance. Positive unit
detuning is one half-linewidth to the red, whereas negative detuning is blue.

The driven PAM4 extinction ratio (ER) compares the highest and lowest optical levels. The
Z-shaped ring reports 3.6~dB outer ER at 100~GBaud under the conditions
in Sec.~\ref{sec:yuantransmission}. Operating closer to the notch can
increase ER while reducing transmitted power and compressing lower-level
spacing. The detuning must therefore provide sufficient outer OMA and
four-level separation, with TDECQ and BER evaluated at the target baud rate.

\subsubsection{Plasma dispersion and tuning efficiency}
Plasma dispersion links changes in the free-carrier density to the
refractive index and absorption of silicon. Reverse bias depletes carriers,
increasing the effective index and reducing free-carrier absorption.
At an electron density of $3\times10^{18}$~cm$^{-3}$ in n-type silicon
and a hole density of $3\times10^{18}$~cm$^{-3}$ in p-type silicon, the
O-band Soref--Bennett fit gives a hole-induced index change relative to
undoped silicon about $1.9\times$ that of electrons
~\cite{soref_jqe1987,reed_np2010,nedeljkovic_pj2011}.
The stronger hole response favors overlap with the p-doped region.
For a uniform segment, optical overlap and active ring fraction weight
the material index change $\Delta n$, giving the resonance shift
\begin{equation}
  \Delta n_\mathrm{eff} = \frac{L_j}{L_\mathrm{rt}}\,\Gamma\,\Delta n,
  \qquad
  \Delta\lambda_\mathrm{res} = \frac{\Delta n_\mathrm{eff}}{n_g}\lambda_\mathrm{res},
  \label{eq:shift}
\end{equation}
where $\Gamma$ is the transverse junction--mode overlap and $L_j$ the active
junction length. The index change $\Delta n_\mathrm{eff}$ is averaged over
the round trip, with segment contributions added. Greater overlap increases
the effective-index change and resonance shift for a given material response.
Operation on the resonance flank converts this shift into intensity
modulation~\cite{xu_oe2007,preble_np2007,li_jstqe2013}.

\begin{figure*}[t]
 \centering
 \includegraphics[width=\textwidth]{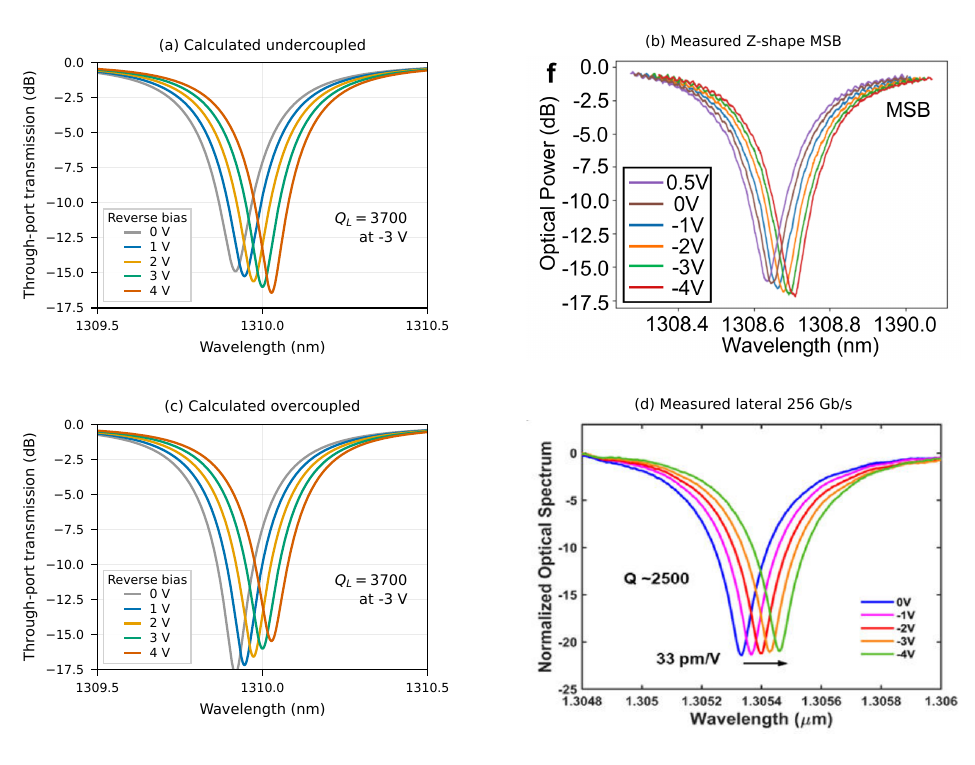}
 \caption{Bias-dependent spectra. \textbf{(a,c)}~Calculated undercoupled
 and overcoupled rings. \textbf{(b,d)}~Measured Z-shaped ring and
 256G lateral ring spectra, not model fits. Cropped/rescaled from Yuan \emph{et al.},
 Fig.~2(f)~\cite{yuan_nc2024},
 \href{https://creativecommons.org/licenses/by/4.0/}{CC BY 4.0}, and
 Xue \emph{et al.}, Fig.~1(c)~\cite{xue_ofc2025},
 \copyright~2025 Optica Publishing Group.}
 \label{fig:biasabsorption}
\end{figure*}

Figure~\ref{fig:biasabsorption} compares calculated spectra for two coupling
regimes with measured spectra from our Z-shaped 5$\times$200~Gb/s
transmitter and the eight-channel transmitter using 256G lateral
rings. Panels (a,c) show the calculated
response as the junction bias varies from 0 to $-4$~V. The two alternative
coupling cases use $Q_L=3700$ and a 16~dB notch at the assumed $-3$~V,
1310~nm reference, with both segments biased together and temperature
and external coupling held fixed. Increasing reverse
bias expands the depletion region, reducing free-carrier absorption within
the optical mode. The intrinsic lifetime $\tau_l$ and quality factor $Q_i$
therefore increase, while $Q_c$ remains approximately fixed. In an
undercoupled ring, $Q_i<Q_c$, the increase in $Q_i$ brings the decay rates
toward balance and deepens the notch until critical coupling is reached.
In an overcoupled ring, $Q_i>Q_c$, the same increase in $Q_i$ moves the decay rates
farther from balance and reduces notch ER. The measured notch deepening
in the Z-shaped ring in panel (b) is consistent with undercoupling over
the displayed reverse-bias range~\cite{yuan_nc2024}. The slight decrease
in ER for the 256G lateral ring in panel (d) is consistent with weak
overcoupling~\cite{xue_ofc2025}.

The tuning efficiency $\eta$ is the resonance shift per volt of increased
reverse bias.
The voltage--length product $V_\pi L$ specifies the voltage and active length
required for a $\pi$ phase shift. Lower values indicate higher efficiency.
The two metrics are related through the free spectral range,
\begin{equation}
  \eta = \frac{\Delta\lambda_\mathrm{res}}{\Delta V}
       = \frac{\mathrm{FSR}}{2\pi}\,\frac{\Delta\phi_\mathrm{rt}}{\Delta V}
       = \frac{\lambda_\mathrm{res}}{n_g}\,\frac{L_j}{L_\mathrm{rt}}\,
         \Gamma\,\frac{\partial n}{\partial V},
  \label{eq:eta}
\end{equation}
\begin{equation}
  V_\pi L = V_\pi L_j = \frac{\mathrm{FSR}\;L_j}{2\,\eta}
          = \frac{\lambda_\mathrm{res}^2\,L_j}{2\,n_g L_\mathrm{rt}\,\eta}
          = \frac{\lambda_\mathrm{res}}{2\,\Gamma\,\partial n/\partial V},
  \label{eq:vpil}
\end{equation}
where $\Delta V$ is the increase in reverse-bias magnitude,
$\Delta\phi_\mathrm{rt}$ the round-trip phase shift and
$\partial n/\partial V$ the material index response per volt.
One FSR corresponds to $2\pi$ of round-trip phase, giving the factor of two
in $V_\pi L$, a figure of merit depending on the
junction cross-section.
The Z-shaped ring's summed tuning efficiency of 27.3~pm/V at 1.6~V swing
per segment gives a 44~pm extreme-state resonance shift. The
256G lateral ring gives a 59~pm shift at 33~pm/V and a 1.8~V swing.
Both demonstrations used an arbitrary-waveform generator and linear
amplifiers~\cite{yuan_nc2024,xue_ofc2025}.
Advanced-node core supplies of about 0.6 to 0.75~V
require voltage-boosting output stages to deliver these swings from a
co-packaged 5-, 3- or 2-nm driver. Junction overlap, doping and active length
must provide the required resonance shift at the delivered voltage.
Greater overlap improves tuning per unit capacitance, whereas increasing junction
capacitance through doping, or increasing the active fraction at fixed ring
length, can also raise tuning efficiency but increases electrical loading and
adds doping-induced absorption to the optical-loss budget.
Stacked output transistors distribute the swing across devices, while
higher-voltage I/O transistors trade speed for voltage tolerance, requiring driver
design to balance voltage sharing, bandwidth and power~\cite{lin_mwcl2022}.
The DC bias and signal extrema, including overshoot, must satisfy junction
leakage, breakdown and transistor terminal-voltage limits across process and
temperature.

\subsubsection{Electro-optic bandwidth}

Junction doping controls depletion width and capacitance. Junction and access
doping, together with contact geometry, determine series resistance $R_s$. For an
abrupt junction with series access paths,
\begin{equation}
  R_s = R_{n,\mathrm{core}} + R_{p,\mathrm{core}}
      + R_{n,\mathrm{slab}} + R_{p,\mathrm{slab}} + R_c + R_m,
  \qquad
  C_j = \frac{\varepsilon_\mathrm{Si}\,h_j L_j}{W}=c_j L_j,
  \label{eq:cjrs}
\end{equation}
where the n- and p-side core and slab resistances, total contact resistance
$R_c$ and internal metal resistance $R_m$ form the device series resistance.
External driver and feed impedances are excluded from $R_s$.
Here $\varepsilon_\mathrm{Si}$ is the silicon permittivity, $h_j$ the effective
junction-wall height, $W$ the depletion width, and $c_j$ the junction
capacitance per unit length. At fixed reverse bias, heavier doping supplies
more ionized dopant charge per unit volume, supporting the junction potential
across a narrower depletion region and increasing capacitance. Increasing
reverse bias expands the depletion region and lowers capacitance. Extending
the active length increases both junction area and optical interaction
length, increasing capacitance and phase shift together.


The electrical bandwidth $f_\mathrm{RC,network}$ is the 3~dB bandwidth of
$V_j/V_\mathrm{src}$ relative to its low-frequency response. When one RC
pole dominates, the lumped estimate $f_\mathrm{RC,lump}\simeq1/(2\pi RC)$
applies, with effective loaded resistance $R$ and capacitance $C$.
The full network retains frequency-dependent driver, feed and shunt impedances
(Sec.~\ref{sec:parasitic}). Distributed or measured S-parameter models are
needed when interconnect propagation and reflections are appreciable.
With a 50~$\Omega$ source, the fitted networks give approximately 79 and
65~GHz for the Z-shaped LSB and MSB segments, respectively
(Table~\ref{tab:ringdesigns}).

Combining loaded electrical and cavity responses gives the approximate
inverse-square EO bandwidth budget,
\begin{equation}
  \frac{1}{f_\mathrm{EO}^{2}}
  \simeq \frac{1}{f_\mathrm{ph}^{2}} + \frac{1}{f_\mathrm{RC,network}^{2}}
  = \left(\frac{\lambda_\mathrm{res} Q_L}{c}\right)^2+\frac{1}{f_\mathrm{RC,network}^{2}},
  \label{eq:bw}
\end{equation}
where $f_\mathrm{EO}$ is an engineering estimate for approximately low-pass
responses, using the loaded electrical bandwidth and photon-lifetime scale.

For the 67~GHz target of the 256G lateral ring, its 92~GHz
photon-lifetime scale requires approximately 98~GHz loaded electrical
bandwidth within this budget, equivalent to a 1.6~ps time constant including
driver and parasitic loading.



\subsubsection{Efficiency--capacitance plane}
\begin{figure}[t]
  \centering
  \includegraphics[width=\textwidth]{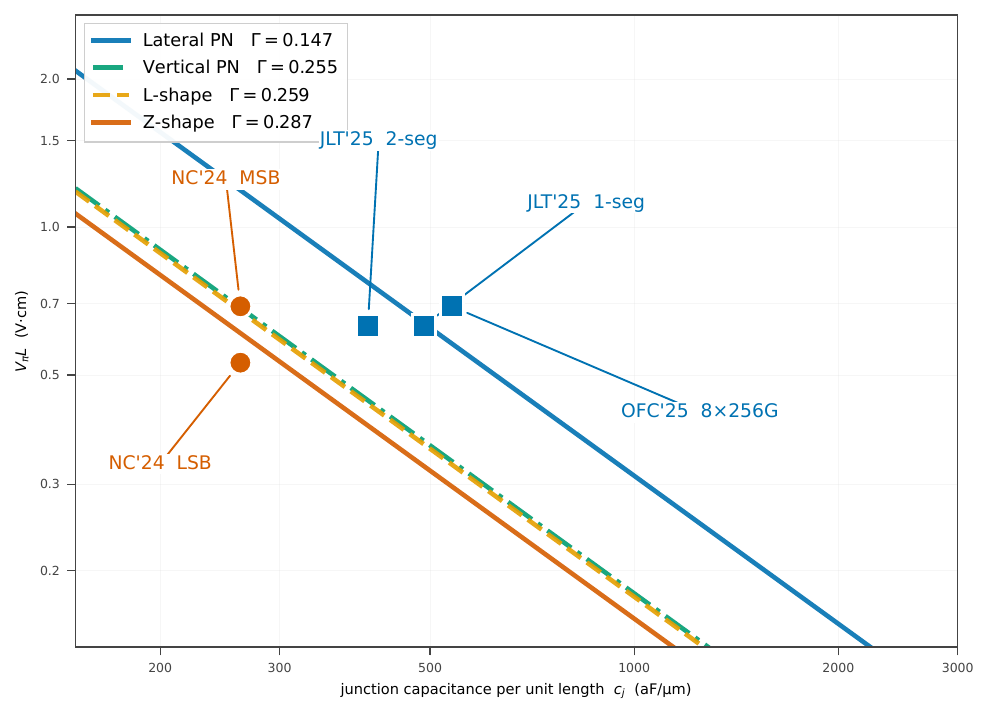}
  \caption{Efficiency--capacitance comparison under Eq.~(\ref{eq:vpilcj}).
  Lines denote fixed-overlap models; markers denote devices.}
  \label{fig:vpilcj}
\end{figure}

Figure~\ref{fig:vpilcj} relates tuning efficiency to electrical loading.
At fixed overlap, a larger capacitance displaces more charge per volt and
reduces the voltage--length product ($V_\pi L$). With carrier density $N$ and
plasma-dispersion coefficient $\partial n/\partial N$, the local
constant-capacitance approximation gives
\begin{equation}
  V_\pi L \;\simeq\;
  \frac{K_0}{\Gamma c_j},
  \qquad
K_0=\frac{\lambda q A_\mathrm{mode}}{2|\partial n/\partial N|}
  \label{eq:vpilcj}
\end{equation}
where $A_\mathrm{mode}$ is the effective optical mode area, $q$ the elementary
charge, and $K_0$ collects the wavelength, area and carrier-response factors.
At a 40~nm depletion width, overlaps of
14.7\%, 25.5\%, 25.9\% and 28.7\% for lateral, vertical,
L-shaped and Z-shaped junctions~\cite{yuan_nc2024} give a $1.9\times$
Z-shaped efficiency advantage over the lateral geometry at equal capacitance
and fixed $K_0$.
The plotted normalization uses nominal one-third/two-thirds round-trip
lengths for the Z-shaped segments and assumed full-active lengths for the
lateral devices. The five markers yield a mean $K_0$ of
45.9~V$\cdot$cm$\cdot$aF/\textmu m with $\pm19\%$ spread. 
This illustrative
coefficient depends on the assumed segment lengths, which were not measured directly.

Under consistent length and bias definitions, $V_\pi L\,c_j$ is independent
of active length, whereas $V_\pi L\,C_j$ scales with total junction length.
The plane compares phase-shifter efficiency at a given capacitance.
Loaded electro-optic bandwidth and transmitter quality additionally depend
on series resistance, driver and pad loading, and the operating point.

Doping changes capacitance and overlap together, requiring a consistent
mode-weighted carrier response in Eq.~(\ref{eq:vpilcj}).
For the 256G lateral ring in the eight-channel transmitter, an n-type concentration of
$10^{19}$~cm$^{-3}$ and a p-type concentration of
$5.3\times10^{18}$~cm$^{-3}$ give a calculated 38.9~nm depletion width
and 20.3~fF junction at $-3$~V.
Its $V_\pi L=0.69$~V$\cdot$cm is close to the Z-shaped value of
0.67~V$\cdot$cm with an n-type concentration of $3\times10^{18}$~cm$^{-3}$
and a p-type concentration of $3\times10^{18}$~cm$^{-3}$ in the 
junction region. The Z-shaped junction thus achieves comparable phase
efficiency at lower doping through greater overlap. 


\subsubsection{Gain--bandwidth product}
At fixed notch depth and normalized detuning, modulation depends on the
resonance shift relative to linewidth. For a voltage swing $\Delta V$,
\begin{equation}
  \Delta\lambda_\mathrm{res}
  \;\simeq\; \frac{\eta\,\Delta V\,Q_L}{\lambda_\mathrm{res}}\Delta\lambda_\mathrm{FWHM}.
  \label{eq:normalized_resonance_swing}
\end{equation}
The multiplier expresses the shift in linewidths. For the Z-shaped ring,
reducing the loaded quality factor $Q_L$ from 3700 to 2500 at 1.6~V per
segment requires summed tuning efficiency to rise from 27 to 40~pm/V
to preserve that shift. The per-volt factor $\eta Q_L/\lambda_\mathrm{res}$
grows with $Q_L$, while photon-lifetime bandwidth decreases inversely,
giving the GBW reference bound within the low-pass approximation of
Eq.~(\ref{eq:bw}),
\begin{equation}
  \mathrm{GBW} \;=\; \frac{\eta\,Q_L}{\lambda_\mathrm{res}}\, f_\mathrm{EO}
  \;\le\; \frac{\eta\,Q_L}{\lambda_\mathrm{res}}\, f_\mathrm{ph}
  \;=\; \frac{\eta\,c}{\lambda_\mathrm{res}^{2}} \;\equiv\; \mathrm{GBW}_\mathrm{max},
  \label{eq:gbw}
\end{equation}

For the 256G lateral ring in the eight-channel transmitter (Table~\ref{tab:ringdesigns}),
33~pm/V at 1310~nm gives a GBW reference of 5.8~GHz/V. The measured
67~GHz bandwidth gives 4.2~GHz/V, or 73\% of that reference, including
electrical loading and detuning-dependent peaking. The modulation-gain GBW
requires bandwidth and low-frequency gain at the same detuning
(Sec.~\ref{sec:ringdyn}).

Finite junction swing also constrains the photon-lifetime bandwidth when
a minimum resonance shift relative to the full linewidth is required.
Combining Eq.~(\ref{eq:normalized_resonance_swing}) with
$f_\mathrm{ph}=c/(\lambda_\mathrm{res} Q_L)$ gives
\begin{equation}
 f_\mathrm{ph}\leq \frac{c\eta\Delta V}{\lambda_\mathrm{res}^2 r_\mathrm{min}},
 \label{eq:depth_bandwidth_bound}
\end{equation}
where $r_\mathrm{min}$ is the required resonance shift divided by the
full linewidth and $\Delta V$ is the delivered junction swing. The bound
assumes fixed tuning efficiency and requires $r_\mathrm{min}$ to come from
an independent target such as extinction ratio or OMA, which also depend on
coupling and detuning. Taking $r_\mathrm{min}$ from the same $Q_L$ makes the
inequality circular and returns the identity
$f_\mathrm{ph}=c/(\lambda_\mathrm{res}Q_L)$. The bound constrains the
photon-lifetime scale alone, whereas the usable electro-optic bandwidth
additionally depends on the loaded electrical network and the peaking of
the detuned cavity (Sec.~\ref{sec:parasitic}).

\subsubsection{Loaded electrical network}\label{sec:parasitic}
The lumped estimate $f_\mathrm{RC,lump}$ describes an isolated junction
charged through $R_s$. In a driven modulator, the pad, substrate and feed
branches also draw current through the source impedance without charging the junction capacitance
$C_j$, and the feed inductance adds a frequency-dependent series term.
An equivalent circuit extracted from measured $S_{11}$ resolves these
branches and gives the loaded transfer function $H_e$ of
Eq.~(\ref{eq:vjvsrc}), the fraction of the source swing delivered to the
junction at each modulation frequency. That fraction sets the electrical
contribution to the electro-optic response and the capacitance optimum of
Fig.~\ref{fig:cjopt}.

\begin{figure*}[t]
 \centering
 \includegraphics[width=\textwidth]{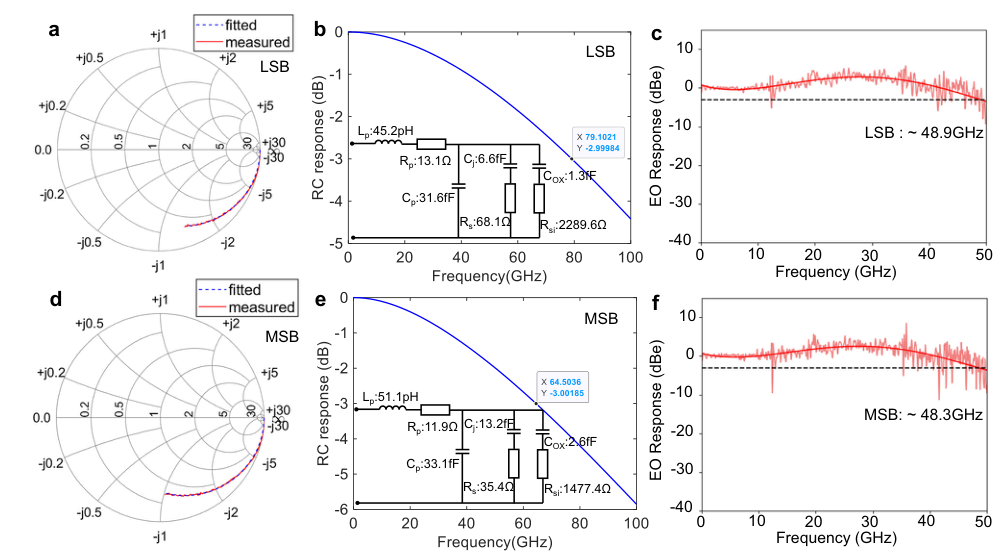}
 \caption{Z-shaped ring characterization at $-3$~V. \textbf{(a,d)}~Measured/fitted
 $S_{11}$. \textbf{(b,e)}~Extracted circuits and calculated responses.
 \textbf{(c,f)}~Measured EO responses. Cropped/rescaled from Yuan
 \emph{et al.}, Fig.~3~\cite{yuan_nc2024},
 \href{https://creativecommons.org/licenses/by/4.0/}{CC BY 4.0}.}
 \label{fig:yuanmeasuredrf}
\end{figure*}

A fitted electrical network for the two-segment Z-shaped ring is shown in
Fig.~\ref{fig:yuanmeasuredrf}. It separates the electrical loading of the
LSB and MSB segments while both segments modulate the same optical cavity.
Fitting the segment $S_{11}$ responses at $-3$~V gives junction
capacitances of 6.6 and 13.2~fF and series resistances of 68.1 and
35.4~$\Omega$, respectively~\cite{yuan_nc2024}. With a 50~$\Omega$ source
and the fitted pad, feed and substrate parasitics, the networks give
junction-voltage bandwidths of 79 and 65~GHz. The $S_{11}$ fit constrains
electrical loading alone. The optical decay and the voltage-to-index
conversion appear only in the electro-optic $S_{21}$, measured as detected
optical power against electrical drive. At 100~pm laser detuning and
$-3$~V bias, the Z-shaped segments have measured EO bandwidths of 49 and
48~GHz [Fig.~\ref{fig:yuanmeasuredrf}(c,f)].
Higher detuning increases bandwidth through optical
peaking while reducing the low-frequency modulation slope.

\begin{figure}[t]
  \centering
  \includegraphics[width=\textwidth]{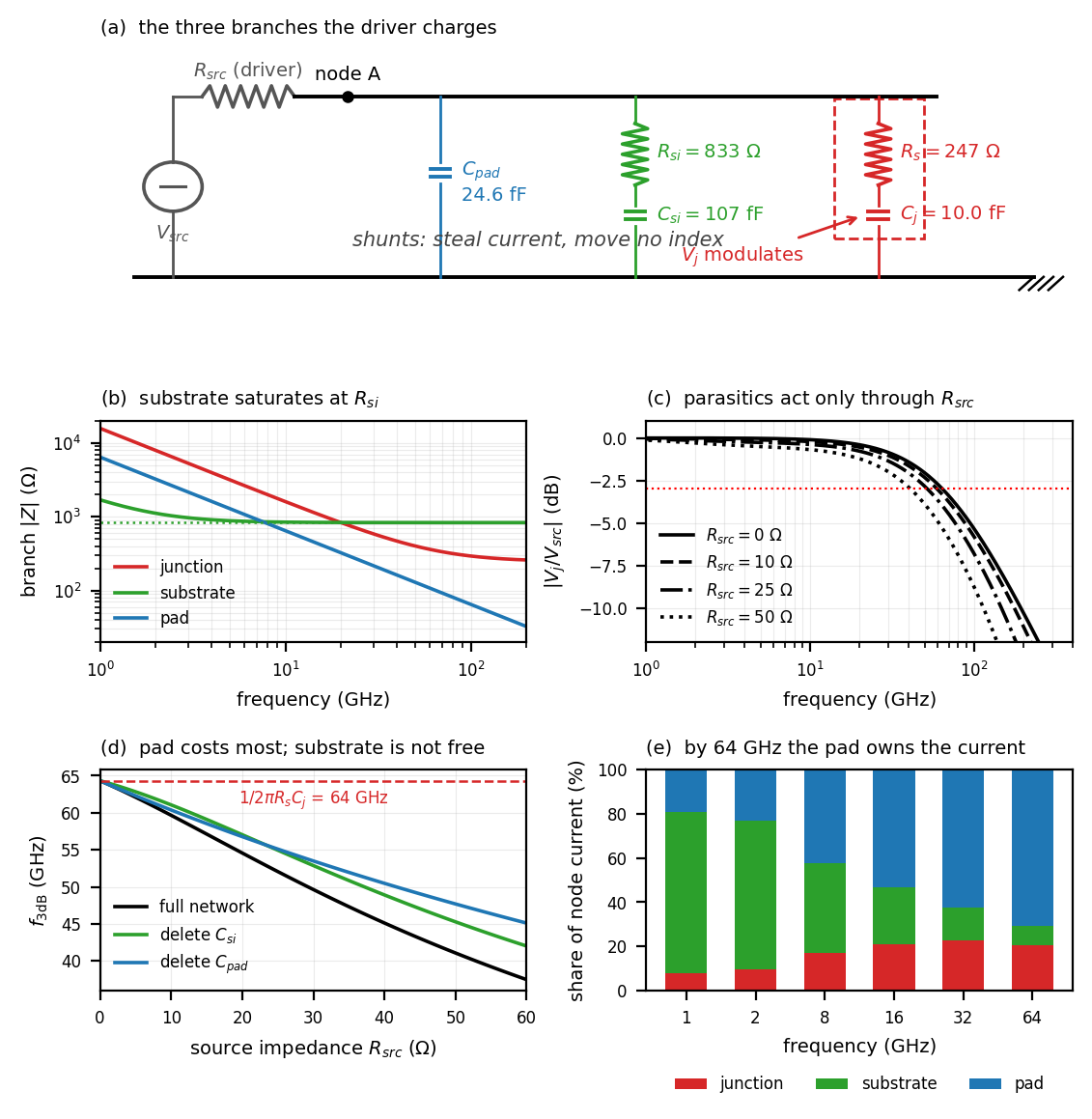}
  \caption{Substrate-resolved lateral loading model~\cite{xue_jlt2025}.
  \textbf{(a)}~Circuit. \textbf{(b)}~Branch impedances.
  \textbf{(c)}~Source-dependent junction response.
  \textbf{(d)}~Bandwidth with individual shunts removed.
  \textbf{(e)}~Branch-current magnitudes normalized by their sum.}
  \label{fig:parasitic}
\end{figure}

Figure~\ref{fig:parasitic} applies Eq.~(\ref{eq:vjvsrc}) to a lateral ring
whose published extraction resolves the substrate
explicitly~\cite{xue_jlt2025}, a different device from the 256G lateral
ring of Fig.~\ref{fig:cjopt}. The junction branch carries
$R_s=246.7$~$\Omega$ and $C_j=10$~fF, the pad $C_\mathrm{pad}=24.6$~fF, and
the substrate branch $C_{si}=107.3$~fF in series with
$R_{si}=833$~$\Omega$. An ideal source gives a junction-voltage bandwidth
of 64~GHz, and a 50~$\Omega$ source 41~GHz. That 23~GHz difference is the
cost of the shunt branches, because current into the pad and substrate
flows through the driver without charging $C_j$. Below its 1.8~GHz
crossover the substrate branch is capacitive, and above it the branch
approaches the resistive load $R_{si}$ [Fig.~\ref{fig:parasitic}(b)], so
its impedance must be retained when evaluating source loading. Xue \emph{et al.} 
report 75~GHz for this segment as retained in
Table~\ref{tab:ringdesigns}. The 64~GHz quoted here is
the intrinsic pole $1/(2\pi R_sC_j)$ computed from their published element
values, so the two numbers rest on different definitions.

Equivalent circuits are not unique. A measured $S_{11}$ constrains the
total input impedance, not the internal arrangement of branches, so a
network without a substrate branch absorbs that current into
$C_\mathrm{pad}$ and $R_s$ and still reproduces the measurement. The
electrical network used by the Z-shaped ring model resolves the feed
resistance and inductance, whereas the substrate-resolved lateral-ring
network instead resolves the substrate branch. Removing each
shunt one at a time and recalculating the junction-voltage bandwidth
reveals how sensitive the overall response is to that particular element,
while comparing normalized branch-current magnitudes shows how the current
divides among the paths [Fig.~\ref{fig:parasitic}(d,e)]. With the branches
separated, the dominant loading element can be identified before a
redesign.

Pad, substrate and junction currents share the feed impedance. Their
total admittance reduces the driven-node voltage, followed by the
intrinsic junction voltage division,
\begin{equation}
\begin{gathered}
  H_e(\omega_m) \equiv \frac{V_j}{V_\mathrm{src}}
  = \underbrace{\frac{1}{1 + Z_\mathrm{feed}Y_A}}_\text{loading}
    \cdot\underbrace{\frac{1}{1 + j\omega_m R_sC_j}}_\text{intrinsic}, \\[2pt]
  Z_\mathrm{feed} = R_\mathrm{src} + R_p + j\omega_m L_p, 
    \qquad
  Y_A = j\omega_m C_\mathrm{pad}
       + \Big(R_{si} + \tfrac{1}{j\omega_m C_{si}}\Big)^{-1}
       + \Big(R_s + \tfrac{1}{j\omega_m C_j}\Big)^{-1},
\end{gathered}
  \label{eq:vjvsrc}
\end{equation}
where $H_e$ is the loaded electrical transfer function,
$\omega_m$ the modulation angular frequency, $Z_\mathrm{feed}$ the
series combination of driver resistance $R_\mathrm{src}$, feed resistance
$R_p$ and inductance $L_p$, and $Y_A$ the driven-node admittance.
Equation~(\ref{eq:vjvsrc}) collects the branches resolved by both
extractions, taking $R_p$ and $L_p$ from the electrical network of the
Z-shaped ring model~\cite{yuan_nc2024} and the substrate branch
$C_{si}$--$R_{si}$ from the substrate-resolved lateral-ring
network~\cite{xue_jlt2025}. Each network is the
special case of Eq.~(\ref{eq:vjvsrc}) with its absent branches set to zero.
The Z-shaped LSB and MSB segments have extracted feed inductances of 45.2 and
51.1~pH, respectively [Fig.~\ref{fig:yuanmeasuredrf}(b,e)].
Their feed resistance and inductance remain at zero external source
resistance. In the substrate-resolved lateral example, $R_p=L_p=0$,
and an ideal source leaves only the intrinsic junction response
[Fig.~\ref{fig:parasitic}(c)].

\subsubsection{Design trade-offs above 200~Gb/s}
\begin{figure}[t]
  \centering
  \includegraphics[width=\textwidth]{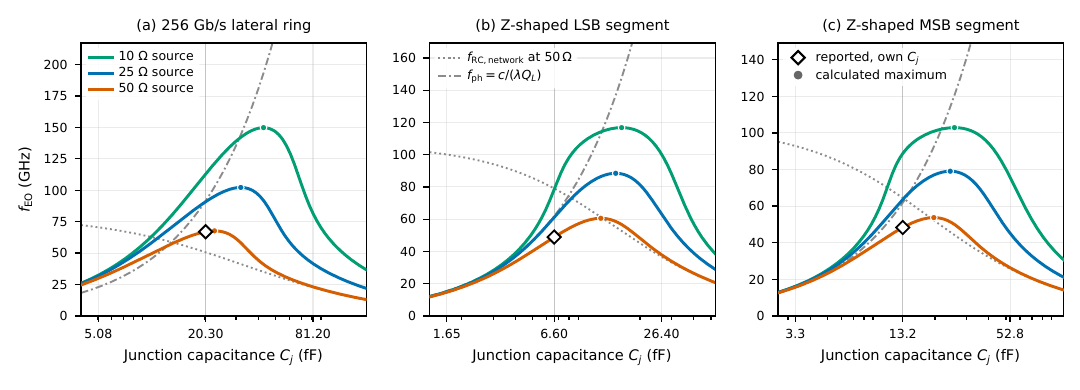}
  \caption{Electro-optic bandwidth against junction capacitance for the
  \textbf{(a)}~256G lateral ring, \textbf{(b)}~Z-shaped LSB and
  \textbf{(c)}~MSB segments. Solid curves are the cascade at 10, 25
  and 50~$\Omega$ source resistance. Grey curves show
  $f_\mathrm{RC,network}$ at 50~$\Omega$ (dotted) and
  $f_\mathrm{ph}$ (dash-dotted). Circles mark calculated maxima and
  diamonds measured bandwidths. Optima are conditional on the stated
  scaling assumptions and fitted detunings.}
  \label{fig:cjopt}
\end{figure}

Higher electro-optic bandwidth requires sufficient resonance tuning while limiting the 
electrical load on the driver. Figure~\ref{fig:cjopt} examines this trade-off to identify 
junction-capacitance targets for higher bandwidth at a given source resistance. 
These calculated targets guide doping and junction-geometry choices through 
Eq.~(\ref{eq:cjrs}), with the associated changes in resistance and optical loss evaluated together.

The comparison separates junction capacitance from parasitic loading. 
The Z-shaped LSB and MSB segments have junction capacitances of 6.6 and 
13.2~fF and fitted pad/via capacitances of 31.6 and 33.1~fF, 
respectively [Fig.~\ref{fig:yuanmeasuredrf}(b,e)]~\cite{yuan_nc2024}. 
For the 256G lateral ring at $-3$~V, the $S_{11}$ fit gives a node 
capacitance near 60~fF~\cite{xue_ofc2025}. Below the reported 167~GHz 
electrical pole, the measured reflection remains predominantly capacitive, 
constraining node capacitance more tightly than series resistance. 
The calculated junction capacitance is 20.3~fF, leaving approximately 
40~fF attributed to pad and routing.

The solid curves in Fig.~\ref{fig:cjopt} combine the loaded electrical 
response $H_e(\omega_m)$ with the cavity response $H_o(\omega_m)$. 
At each operating point, $H_o$ describes small-signal resonance modulation 
with intrinsic loss and coupling held fixed and is normalized to unity at zero 
frequency. The combined response is
\begin{equation}
  H_\mathrm{EO}(\omega_m)
    = H_e(\omega_m)\;H_o(\omega_m),
  \label{eq:eotf}
\end{equation}
and $f_\mathrm{EO}$ is the first frequency at which $|H_\mathrm{EO}|^2$ falls 
3~dB below its DC value. An effective detuning is fitted to each reference 
electro-optic bandwidth at a 50~$\Omega$ source resistance, with the electrical 
parameters held fixed. This calibration retains cavity peaking, which is omitted 
from the approximate low-pass budget of Eq.~(\ref{eq:bw}).

Figure~\ref{fig:cjopt} uses total junction capacitance $C_j$ because 
source loading depends on the individual junction and parasitic branches. 
The Z-shaped LSB and MSB segments share a cross-section and capacitance 
per unit length, but their total junction capacitances differ by a factor 
of two. Their separate electrical networks must therefore be retained when 
comparing capacitance optima. Under inverse-length resistance scaling, the 
intrinsic product $R_sC_j=\rho_sc_j$ is length-independent, where $\rho_s=R_sL_j$.

The capacitance sweep assumes fixed overlap and active length. Increasing $C_j$ 
then increases the charge displaced per volt and improves tuning efficiency. 
To preserve the resonance shift relative to linewidth at the same junction swing, 
the cavity is redesigned with a lower loaded quality factor. This broadens 
the photon linewidth while the increased capacitance slows junction charging. 
Series resistance, pad, feed and substrate parameters remain fixed during the sweep.

At fixed wavelength, geometry, overlap $\Gamma$ and carrier-response factor $K_0$, 
the product $(V_\pi L)c_j\simeq K_0/\Gamma$ is constant, 
giving $\eta=\mathrm{FSR}\,L_j/[2(V_\pi L)]\propto c_j$. Maintaining the same 
relative resonance shift and junction swing $\Delta V$ gives the following 
scalings, with junction-dominated electrical loading and fixed $\rho_s$,
\begin{equation}
 \begin{aligned}
  Q_L &= \frac{m\lambda_\mathrm{res}}{\eta\Delta V}
      \;\propto\; c_j^{-1},
  \qquad
  f_\mathrm{ph} &= \frac{1}{2\pi\tau_p}
     = \frac{c}{\lambda_\mathrm{res} Q_L} \;\propto\; c_j,
  \qquad
    f_\mathrm{RC,lump} \;\propto\; c_j^{-1}.
 \end{aligned}
  \label{eq:cjopt}
\end{equation}
where $m$ is the resonance shift divided by the full linewidth. These assumptions make the calculated optima conditional design targets. A doping change must also be evaluated for its effects on resistance, optical overlap and absorption.

For the assumed active lengths of 39.5~$\mu$m for the Z-shaped ring and 37.7~$\mu$m 
for the 256G lateral ring, total capacitance follows $C_j=c_jL_j$. Increasing source 
resistance from 10 to 50~$\Omega$ shifts the optimum capacitance 
from 25.7 to 19.7~fF for the Z-shaped ring and from 42.9 to 22.8~fF for the 256G lateral ring. 
The corresponding peak bandwidths fall from 103 to 54~GHz and from 150 to 68~GHz. 
The estimated 20.3~fF junction of the 256G lateral ring is already near its 50~$\Omega$ optimum. 
Within this model, reducing source resistance offers more bandwidth improvement than further 
reducing junction capacitance.

\begin{figure*}[t]
  \centering
  \includegraphics[width=\textwidth]{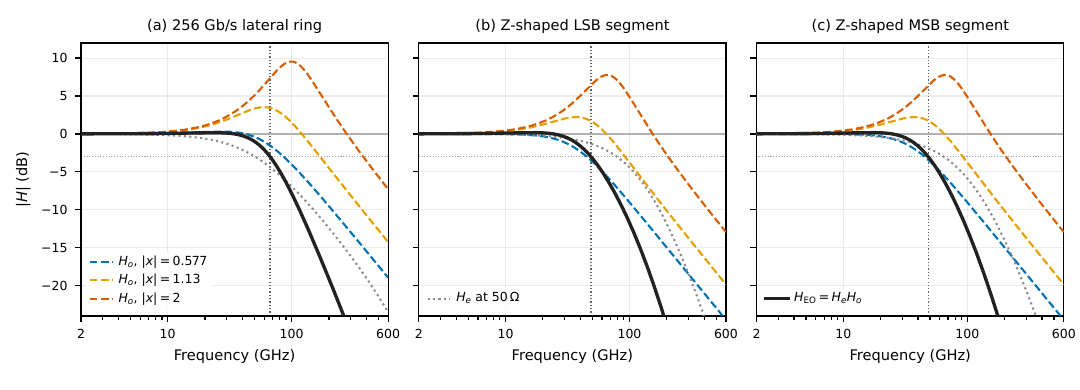}
  \caption{Frequency responses for the \textbf{(a)}~256G lateral ring,
  \textbf{(b)}~Z-shaped LSB and \textbf{(c)}~MSB segments.
  Dashed curves show $H_o$ at three detunings, dotted curves show $H_e$
  at 50~$\Omega$, and solid curves show $H_\mathrm{EO}$ at the fitted
  detuning and the same source resistance. Vertical lines mark measured
  electro-optic bandwidths.}
  \label{fig:eoresponse}
\end{figure*}

Figure~\ref{fig:eoresponse} separates the electrical and optical contributions in 
the frequency domain. For the 256G lateral ring [Fig.~\ref{fig:eoresponse}(a)], 
the loaded electrical cutoff is 51~GHz (dotted curve), yet the composite electro-optic 
response reaches 67~GHz (solid curve). This bandwidth extension occurs because optical 
cavity peaking compensates for the high-frequency electrical roll-off. Modeled cavity 
peaking increases to approximately 3.5~dB at $|x|=1.13$ and 9.6~dB at $|x|=2$ (dashed curves). 
At the fitted operating detuning ($|x|=0.90$), approximately 2~dB of optical peaking near 
48~GHz acts as an analog pre-emphasis filter that lifts the attenuated electrical signal, 
flattening the cascaded response $H_\mathrm{EO} = H_e H_o$ and extending the net 3~dB 
bandwidth to 67~GHz with less than 0.2~dB of passband ripple.

This detuning of $|x|=0.90$ is a calibrated fit selected to match the measured 67~GHz 
bandwidth under the published zero-inductance lumped network ($L_p=0$), rather than a 
directly measured operating point. Xue \emph{et al.} report operating at the 3~dB 
through-port insertion loss point, corresponding to $|x|\approx0.98$, which yields a calculated 
bandwidth of 72.5~GHz under this idealized lumped model. If on-chip routing instead 
introduces a modest parasitic inductance of $L_p=20$~pH, the loaded electrical cutoff rises 
from 51 to 58~GHz, so reaching 67~GHz requires less cavity peaking and shifts the matching 
detuning inward to $|x|\approx0.80$, near the maximum dynamic OMA point. These variations 
illustrate that extracting junction dynamics and predicting composite bandwidth require 
precise knowledge of the laser parking offset relative to the hot resonance.

For the Z-shaped ring [Fig.~\ref{fig:eoresponse}(b,c)], the loaded electrical cutoffs 
are 79~GHz for the LSB segment and 65~GHz for the MSB segment. Because the Z-shaped 
device operates closer to the notch ($|x| \approx 0.58\text{--}0.62$) to maximize optical 
modulation amplitude, cavity peaking remains below 1~dB, leaving the electro-optic 
bandwidths (49 and 48~GHz) below the electrical cutoffs. The solid electro-optic responses 
use the same networks and fitted detunings as Fig.~\ref{fig:cjopt}, including the parasitics 
extracted for the Z-shaped ring and the assumed junction/pad partition of the 256G lateral ring.

\subsection{Optical nonlinearities in the cavity}\label{sec:selfheat}
Absorption of the circulating optical field heats the waveguide and increases its effective index. The resulting resonance redshift changes the laser detuning, modulation slope and OMA~\cite{decea_oe2019}. Thermal control must therefore maintain the operating point as optical power and temperature vary. At fixed bias, a small temperature change gives
\begin{equation}
 \Delta\lambda_\mathrm{res}=\frac{\lambda_\mathrm{res}}{n_g}
                  \frac{\partial n_\mathrm{eff}}{\partial T}\Delta T.
 \label{eq:thermoopticshift}
\end{equation}
where $\Delta T$ is the effective temperature change sampled by the optical mode, with thermal expansion neglected.

At thermal equilibrium, the optical-heating contribution follows the average absorbed power~\cite{bao_jlt2024},
\begin{equation}
 \overline{P}_\mathrm{abs}
 =\left\langle P_\mathrm{abs}(t)\right\rangle,
 \qquad \Delta\lambda_\mathrm{res}
     =\left.\frac{\Delta\lambda_\mathrm{res}}{\overline{P}_\mathrm{abs}}
  \right|_\mathrm{ref}
            \overline{P}_\mathrm{abs},
 \label{eq:opticalheat}
\end{equation}
where $P_\mathrm{abs}$ is the thermalized optical power. The reference ratio is evaluated at nonzero absorbed power, with the resonance shift measured from zero optical heating at fixed bias, heater power and package temperature.

For continuous-wave (CW) input at fixed bus power, intrinsic loss and external coupling, constant thermal sensitivity gives
\begin{equation}
  \Delta\lambda_\mathrm{res}
    =\frac{\left.\Delta\lambda_\mathrm{res}\right|_{x=0}}{1+x^2}.
 \label{eq:thermal_spectrum}
\end{equation}
Here $\left.\Delta\lambda_\mathrm{res}\right|_{x=0}$ is the on-resonance thermal shift at the same bus power. The detuning $x$ is measured from the hot resonance, whose position depends on absorption. Absorption and resonance shift must therefore be solved self-consistently. Through-port transmission follows Eq.~(\ref{eq:xnorm}), including off-resonance loss. An assumed peak absorbed fraction of approximately 84\% and thermal sensitivity of 160~pm per milliwatt absorbed give an on-resonance shift of 134~pm per milliwatt of bus power. Heater tuning~\cite{xue_ofc2025} enters the combined wavelength balance in Sec.~\ref{sec:coldplace}.

For a blue-detuned laser ($\lambda<\lambda_\mathrm{res}$), heating moves the resonance farther from the laser, reducing absorption and opposing further heating. With red detuning, the resonance initially moves toward the laser, increasing absorption and reinforcing heating. This feedback distorts the transmission recorded during a slow laser-wavelength scan and can produce multiple stable thermal states. In the model of Sec.~\ref{sec:va}, absorbed power drives a thermal wavelength-shift state that updates the cavity detuning. The spectra and eyes in Fig.~\ref{fig:hoter} describe modulation about the settled hot resonance.

Two-photon absorption also generates electron--hole pairs at a rate proportional to the squared local optical intensity. The accumulated carriers lower the effective index through free-carrier dispersion (FCD), shifting the resonance blue through Eq.~(\ref{eq:shift}). Free-carrier absorption shortens the intrinsic lifetime and changes the notch depth and linewidth. Reverse bias sweeps carriers out, reducing their residence time and both effects~\cite{decea_oe2019}. The Z-shaped ring combines this carrier extraction with two-segment modulation.

Photocurrent can additionally change the operating bias. Current through the DC feed resistance produces a voltage drop that reduces the junction's reverse bias. The depletion region contracts, increasing the carrier population sampled by the optical mode and producing a further blue shift. With 33~pm/V tuning, 100~\textmu A through 1~k$\Omega$ gives approximately 3~pm of shift. Low-resistance feeds help preserve junction bias and the relative modulation weights of the Z-shaped segments.

\begin{figure}[t]
 \centering
 \includegraphics[width=\textwidth]{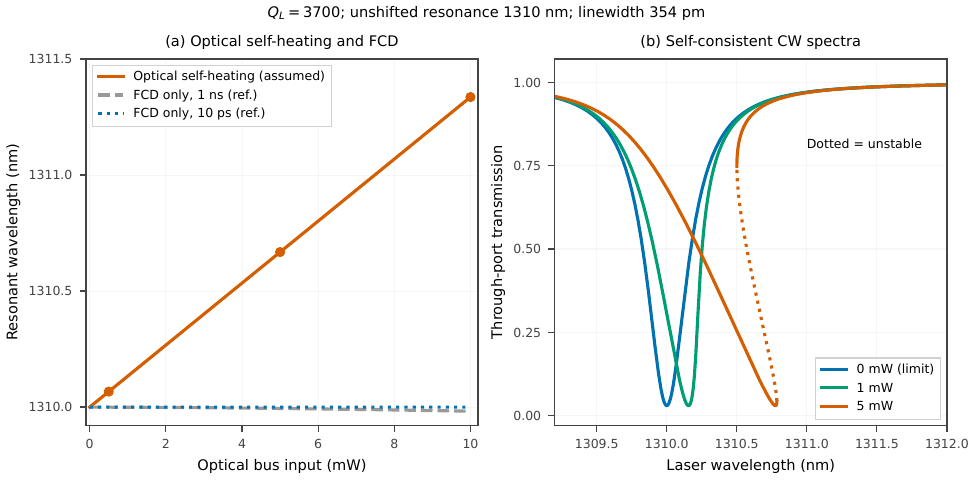}
 \caption{Calculated power-dependent response at $Q_L=3700$.
 \textbf{(a)}~Separate thermal-only and FCD-only resonance wavelengths versus bus power.
 \textbf{(b)}~CW transmission equilibria versus laser wavelength, stable (solid) and unstable (dotted).
 The 0~mW curve is the weak-probe limit.}
 \label{fig:thermalmechanisms}
\end{figure}

Figure~\ref{fig:thermalmechanisms}(a) compares separate thermal and carrier contributions using a 1310~nm reference and $Q_L=3700$, corresponding to the Z-shaped ring's 354~pm linewidth. The assumed thermal shift reaches one linewidth at 2.7~mW bus input. The carrier curves use reference carrier-density and mode-volume parameters from the 256G lateral-ring model, with residence times of 10~ps with sweep-out and 1~ns without it. These curves illustrate the individual mechanisms rather than a coupled thermal--carrier solution for the Z-shaped ring.

At 5~mW bus input, the assumed thermal redshift is 668~pm, compared with reference FCD blue shifts of 5.7~pm at 1~ns residence time and 0.1~pm at 10~ps. Faster carrier extraction reduces the blue shift, while heating dominates the resonance displacement under these assumptions. The thermal shift therefore provides the main contribution to the required wavelength-compensation range in this comparison.

Panel~(b) shows calculated CW through-port transmission versus laser wavelength at fixed bus powers of 0, 1 and 5~mW. The model retains $Q_L=3700$ and the same thermal sensitivity, but sets both peak absorption and notch depth to 97\%, with an unheated resonance at 1310~nm. The 0~mW curve denotes the weak-probe limit. The horizontal axis is the scanned laser wavelength, while the ring resonance changes with the self-consistent thermal state.

Above the calculated 1.8~mW bistability threshold, the equilibrium transmission curve folds back over a range of laser wavelengths. Within that range, one laser wavelength can support two stable hot-resonance positions separated by an unstable state. During a slow scan, transmission follows a stable branch until that branch ends, then jumps to another branch. The jump occurs at different wavelengths for opposite scan directions, producing hysteresis. Laser and heater acquisition must therefore reach the intended operating branch.

The threshold scales with linewidth and inversely with peak absorbed fraction and thermal sensitivity. Reducing $Q_L$ from 3700 to 2500 raises the calculated threshold to 2.6~mW when the other assumptions are unchanged. This threshold marks the onset of possible hysteresis during acquisition. Blue-detuned equilibria remain locally stable above it in the thermal-only model because heating reduces absorption.

The operating-power example in Sec.~\ref{sec:operatingpower} uses 11.3~mW bus input for the 256G lateral ring. At the selected hot detuning, approximately 5.8~mW is absorbed, giving a 920~pm redshift spanning 1.8 linewidths. Cold resonance placement and heater reserve are chosen together to maintain that operating point. The 15~mW bus-power cap in Sec.~\ref{sec:ratebudget} is an analysis assumption. Establishing a device power limit additionally requires checking tuning range, photocurrent-induced bias changes, nonlinear absorption, dynamic stability and eye quality at the intended operating conditions.

\subsection{Laser detuning and the operating point}\label{sec:ringdyn}

\subsubsection{Laser detuning}
Laser detuning determines how efficiently a resonance shift produces optical modulation and how the cavity responds to rapid voltage changes. The operating point must therefore balance modulation amplitude, bandwidth and waveform distortion.

In the depletion-width model of Fig.~\ref{fig:hoter}, increasing reverse bias expands the depletion region, raising the effective index and reducing absorption. At fixed laser wavelength, the voltage-dependent resonance position and linewidth determine the detuning through Eq.~(\ref{eq:xnorm}). The detuning and notch depth at each voltage level then determine the settled optical power. Integrating Eq.~(\ref{eq:cmt_mrm}) with the filtered junction voltage includes cavity memory and interference with the directly transmitted field (Sec.~\ref{sec:va}). Beyond the maximum-slope operating point, increasing detuning can broaden the optical response through peaking while reducing low-frequency modulation gain.

At 1310~nm, unit normalized detuning corresponds to half a linewidth, or 177~pm for the Z-shaped ring at $Q_L=3700$ and 262~pm for the 256G lateral ring at $Q_L=2500$. Their respective extreme-state resonance shifts of 44 and 59~pm span approximately 0.25 and 0.23 in normalized detuning [Fig.~\ref{fig:boostbias}].

\subsubsection{Maximum slope, optical modulation amplitude and gain--bandwidth}\label{sec:ratebudget}
Maximum slope, maximum optical modulation amplitude (OMA) and maximum gain--bandwidth product (GBW) describe different operating objectives. For a fixed-depth Lorentzian, maximum slope occurs at $|x|\simeq0.58$, where zero local curvature suppresses second-order static distortion~\cite{yu_oe2014}. The static finite-swing OMA, defined by the outer settled powers, peaks nearby for the fixed-loss 256G lateral-ring model at $Q_L=2500$ and 1.8~V$_\mathrm{pp}$. At 128~GBaud, cavity dynamics move the waveform-extrema OMA maximum to $|x|\simeq0.80$ [Table~\ref{tab:biascriteria}]. Waveform extrema include transition overshoot, whereas the sampled eye opening depends on adjacent-level separation at the decision instant.

The detuning-dependent GBW combines low-frequency modulation gain with the 3~dB bandwidth~\cite{yu_oe2014,karimelahi_oe2016}. This definition differs from the resonance-shift metric in Eq.~(\ref{eq:gbw}). Electrical filtering reduces the high-frequency enhancement, while further detuning lowers the low-frequency gain. With the assumed 167~GHz electrical pole, the grid-sampled GBW optimum shifts from the bare-cavity value $|x|=1.14$ to $|x|=0.93$ [Table~\ref{tab:biascriteria}].

\begin{figure*}[t]
 \centering
 \includegraphics[width=\textwidth]{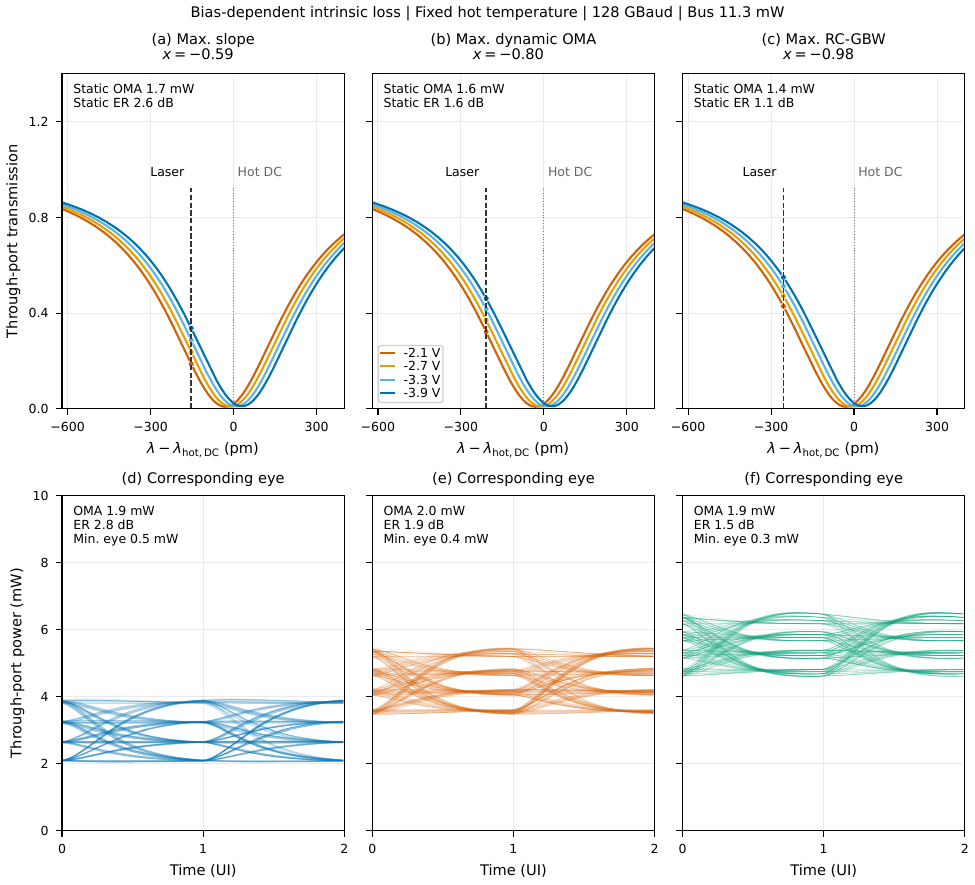}
 \caption{Calculated spectra \textbf{(a--c)} and noise-free PAM4 eyes
 \textbf{(d--f)} at maximum slope, waveform OMA and RC-inclusive GBW,
 respectively. Bias-dependent loss, 128~GBaud, 11.3~mW bus input,
 1.8~V$_\mathrm{pp}$ and a 167~GHz drive pole are used.
 Dashed lines mark the laser; markers denote settled levels.}
 \label{fig:hoter}
 \label{fig:eye}
\end{figure*}

Voltage-dependent absorption changes these optima because the linewidth and notch depth vary during modulation. With fixed external coupling, the loss-aware 256G lateral-ring model gives a notch extinction ratio (ER) of 19.6--20.5~dB across the drive swing [Fig.~\ref{fig:hoter}]. The maximum-slope, waveform-OMA and GBW operating points occur at $x=-0.59$, $-0.80$ and $-0.98$, respectively. The reference model uses a 524~pm linewidth, coupling ratio $\rho=1.2$ at $-3$~V and tuning efficiency of 33~pm/V. An n-type concentration of $10^{19}$~cm$^{-3}$, a p-type concentration of $5.3\times10^{18}$~cm$^{-3}$ and an assumed 350~nm absorbing width determine the undepleted-region loss. Of the reference 151~dB/cm loss, 31~dB/cm remains fixed.

\begin{table}[t]
\centering
\caption{Fixed-loss bias comparison at $Q_L=2500$, 128~GBaud and
1.8~V$_\mathrm{pp}$.}
\label{tab:biascriteria}
\small
\setlength{\tabcolsep}{4pt}
\begin{tabular}{lccccc}
\toprule
\textbf{Objective} & $|x|$ & \textbf{Offset} & \textbf{Mean IL}
 & $k_\mathrm{OMA}$ & $P_\mathrm{bus,req}$ \\
 & & (pm) & (dB) & (\%) & (mW) \\
\midrule
Local slope & 0.58 & 151 & 5.9 & 16 & 11.9 \\
Static finite-swing OMA & 0.58 & 153 & 5.8 & 16 & 11.9 \\
Dynamic waveform OMA & $\sim0.80$ & 210 & 4.0 & 17 & 11.3 \\
GBW with 167~GHz RC & $\sim0.93$ & 244 & 3.3 & 17 & 11.5 \\
Bare-cavity GBW & $\sim1.14$ & 299 & 2.5 & 16 & 12.1 \\
\bottomrule
\end{tabular}
\end{table}

The conversion factor $k_\mathrm{OMA}$ is the waveform-extrema OMA divided by ring-bus input power. Mean insertion loss (IL) is calculated from the mean linear power of the same record, excluding coupling and routing losses. The powers in Table~\ref{tab:biascriteria} assume 0.6~mW receiver OMA and 5~dB downstream loss (Sec.~\ref{sec:operatingpower}). Compared with maximum waveform OMA, the RC-inclusive GBW operating point reduces mean insertion loss by 0.7~dB but also reduces modulation contrast, leaving a 0.063~dB increase in required power. Under the continuous-wave (CW) assumptions of Sec.~\ref{sec:selfheat}, absorbed power falls from 5.8 to 5.1~mW, reducing thermal redshift.

\begin{figure*}[t]
	\centering
	\includegraphics[width=\textwidth]{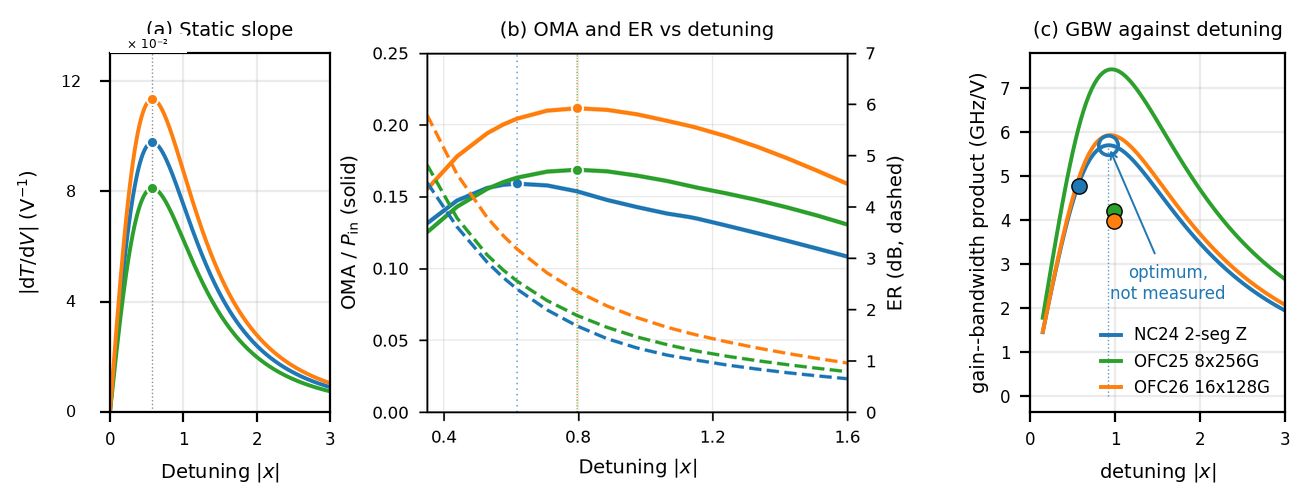}
	\caption{Detuning-dependent response. \textbf{(a)}~Static modulation slope.
	\textbf{(b)}~Simulated waveform OMA (solid) and ER (dashed).
	\textbf{(c)}~GBW from measured electro-optic bandwidths (filled markers) and the
	predicted Z-shaped ring optimum (open). Device-specific rates and drives
	are specified in the text.}
	\label{fig:boostbias}
\end{figure*}

For our Z-shaped ring, the overcoupled approximation in Fig.~\ref{fig:boostbias}(a,c) uses the Z-shaped LSB segment's 79~GHz electrical pole and a combined tuning efficiency of 27~pm/V, with both segments perturbed together. The calculated GBW maximum is near $|x|=0.92$. At 100~GBaud and 1.6~V$_\mathrm{pp}$ per non-return-to-zero (NRZ) segment, waveform OMA peaks near $|x|=0.62$ and reaches 16\% of bus input power, using an assumed 9/18~pm/V tuning split [Fig.~\ref{fig:boostbias}(b)]. The undercoupled model used for the noise comparison retains both fitted segment networks and gives a GBW optimum near $|x|=0.97$ (Sec.~\ref{sec:va}).

For the 128~Gb/s lateral ring~\cite{bu_ofc2026}, the model gives a normalized waveform OMA of 21\% near $|x|=0.80$ at 64~GBaud. The calculation uses $Q_L=4300$, 1.6~V$_\mathrm{pp}$, 27~pm/V tuning, an 18~dB overcoupled notch and an estimated 84~GHz electrical pole. This larger OMA is obtained at a lower baud rate and with a narrower resonance than in the 256G lateral-ring model. All waveform OMA and ER values in Fig.~\ref{fig:boostbias}(b) use noise-free extrema before receiver processing. Power normalization in the plots and Table~\ref{tab:eyequality} refers to the ring bus after input-coupling loss.

\begin{figure*}[t]
 \centering
 \includegraphics[width=\textwidth]{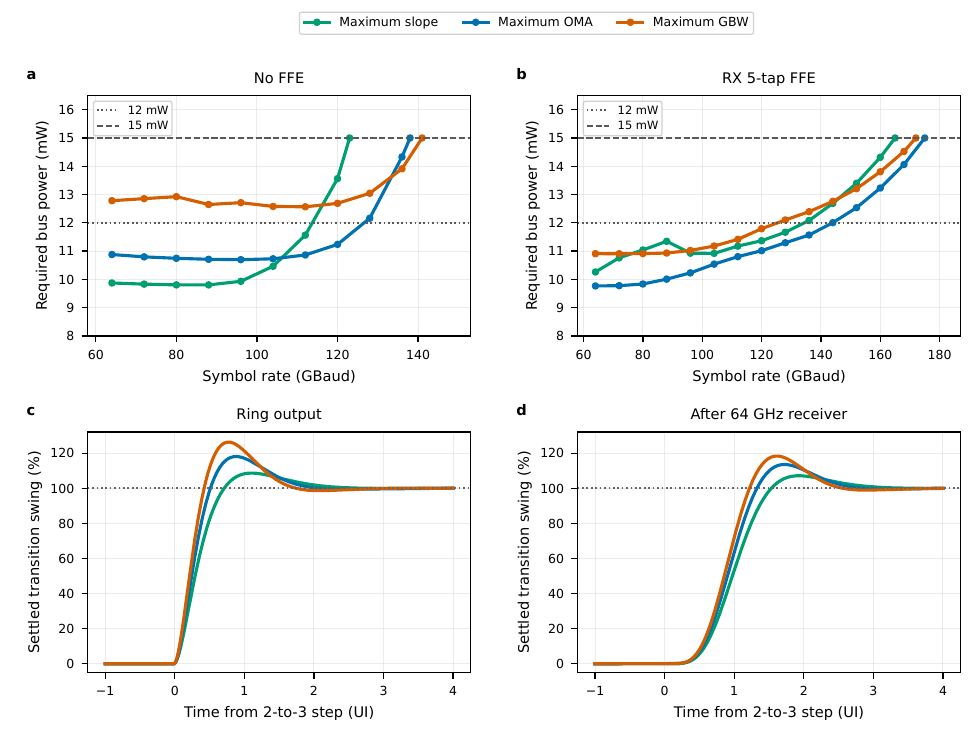}
 \caption{Fixed-loss operating-point comparison.
 \textbf{(a,b)}~Required bus power versus symbol rate without and with a
 five-tap receiver feed-forward equalizer; each trace stops at the highest rate
 meeting the $4.8\times10^{-4}$ symbol-error target within 15~mW. Dotted and
 dashed lines mark 12 and 15~mW. \textbf{(c,d)}~Ring and receiver 2-to-3 steps
 at 128~GBaud. Transmitter equalization is bypassed and the receiver is held at
 64~GHz.}
 \label{fig:operatingrate}
\end{figure*}

Receiver filtering and equalization determine whether these operating-point advantages translate into lower required power or a higher supported baud rate. Figure~\ref{fig:operatingrate}(a,b) compares the operating points of the fixed-loss 256G lateral-ring model, retaining $Q_L=2500$, coupling ratio $\rho=1.22$, 33~pm/V tuning and 1.8~V$_\mathrm{pp}$. This comparison uses the refined GBW bias $x=-0.96$. The electrical pole and Bessel receiver bandwidth remain at 167 and 64~GHz as baud rate increases, and transmitter feed-forward equalization (FFE) is bypassed.

The sampling phase, decision thresholds and optional receiver equalizer are determined from training data. Gaussian noise is then applied statistically to held-out decision samples to calculate the symbol error rate (SER),
\begin{equation}
 \begin{aligned}
 p_{e,i}(\sigma)&=\mathcal{Q}\!\left(\frac{y_i-b_{\ell_i}}{\sigma}\right)
	 +\mathcal{Q}\!\left(\frac{b_{\ell_i+1}-y_i}{\sigma}\right),
\quad
 \mathrm{SER}(\sigma)&=\frac{1}{4}\sum_{\ell=0}^{3}
 \frac{1}{N_\ell}\sum_{i:\,\ell_i=\ell}p_{e,i}(\sigma),
 \end{aligned}
 \label{eq:sample_ser}
\end{equation}
where $y_i$ is the noise-free decision sample, $\ell_i$ its transmitted level, $b_\ell$ the decision boundary, $\sigma$ the root-mean-square (RMS) noise amplitude and $N_\ell$ the number of samples at level $\ell$. The function $\mathcal{Q}$ is the standard-normal upper tail, with $b_0=-\infty$ and $b_4=+\infty$. Errors are averaged with equal weight for the four transmitted levels. The target is $1.5\mathcal{Q}(3.41)\simeq4.8\times10^{-4}$.

The propagated relative intensity noise (RIN) variance is subtracted from the total variance allowed at the target SER. The positive remainder determines the additive RMS noise margin $M(P_0)$, referred to the input of the receiver equalizer with downstream attenuation factored out. At fixed detuning, linear optical-power scaling gives the required bus and laser powers,
\begin{equation}
 \begin{aligned}
 P_\mathrm{bus,req}&=P_0\frac{\sigma_\mathrm{add}}
	  {10^{-L_\mathrm{down}/10}M(P_0)},
  \quad
 P_\mathrm{laser,req}&=P_\mathrm{bus,req}\,10^{L_\mathrm{in}/10},
 \end{aligned}
 \label{eq:rate_required_power}
\end{equation}
where $P_0$ is the reference bus power, $\sigma_\mathrm{add}$ the fixed optical-power-equivalent receiver noise before equalization, and $L_\mathrm{down}$ and $L_\mathrm{in}$ the downstream and input losses in dB.

The additive noise of 21.4~$\mu$W RMS is calibrated at the maximum-OMA bias with receiver FFE, 11.3~mW bus power and 128~GBaud. The calculation assumes 5~dB downstream loss, $-155$~dB/Hz input RIN over 0--64~GHz and 2.5~dB input loss, with no additional margin. Shot noise, jitter, dispersion and forward-error correction are excluded. Residual intersymbol interference (ISI) is measured as the RMS spread about each level mean, normalized by the outer-level centroid separation. Power and ISI comparisons use 128~GBaud. The highest supported rates satisfy the SER target on both tested symbol records within an assumed 15~mW bus-power cap.

For the reference record at 128~GBaud, the maximum-slope, OMA and GBW operating points require 17.5, 12.2 and 13.0~mW without receiver equalization, and 11.7, 11.3 and 12.1~mW with five-tap receiver FFE, respectively. Maximum OMA requires the least power in both cases. The highest supported rate has a different ranking. Without equalization, the lower ISI at the GBW operating point supports the highest rate within the 15~mW cap. Receiver FFE reverses this ranking, allowing the maximum-OMA operating point to support the highest rate [Fig.~\ref{fig:operatingrate}(a,b)].

Maximum slope produces the least overshoot [Fig.~\ref{fig:operatingrate}(c,d)] but the most ISI before equalization. After equalization, its smaller RMS level spread still gives less noise margin than the maximum-OMA operating point because the absolute separation between levels is also smaller. Operating-point selection therefore requires both residual distortion and optical level separation to be evaluated with the intended receiver.

\subsubsection{Chirp}
Voltage-induced resonance shifts change both the through-port optical power and phase in Eq.~(\ref{eq:cmt_mrm}). The time derivative of phase relative to the laser gives the instantaneous frequency excursion, or chirp. For the fixed-loss 256G lateral-ring model at $Q_L=2500$, 1310~nm, 128~GBaud, 1.8~V$_\mathrm{pp}$ and hot detuning $x=-0.80$, the calculated RMS chirp is 1500~MHz.

Chirp affects the detected intensity when the channel introduces frequency-dependent delay or unequal sideband transmission. A chirped waveform and a constant-phase control with identical transmitted intensity produce the same detected waveform through a flat channel. Their RMS difference increases to 1\% after an assumed integrated dispersion of 2.5~ps/nm and to 3\% after an assumed single-pole optical bandpass filter with a 120~GHz power full width at half maximum, centered 30~GHz above the laser. Both comparisons use a periodic 256-symbol PAM4 record, identical timing and a fourth-order 64~GHz Bessel receiver without equalization or added noise. The differences are normalized to the flat-channel detected peak-to-peak excursion.

\section{Thermal tuning and resonance placement}\label{sec:mrmnl}

At 128~GBaud PAM4, the selected $x=-0.80$ bias minimizes modeled bus
power among the fixed-loss candidates at equal receiver OMA
(Sec.~\ref{sec:ringdyn}), placing the laser 210~pm blue of the 256G lateral
ring's hot resonance. Self-heating changes the alignment after turn-on,
coupling optical power to cold resonance placement and heater control.

\subsection{Operating optical power}\label{sec:operatingpower}
Using the waveform-extrema conversion $k_\mathrm{OMA}$ defined in
Sec.~\ref{sec:ringdyn}, the required ring-input bus power is
\begin{equation}
 P_{\mathrm{bus,dBm}}=S_\mathrm{RX}+L_\mathrm{post}+M
                      -10\log_{10}k_\mathrm{OMA},
 \label{eq:operatingpower}
\end{equation}
where $S_\mathrm{RX}$ is the required receiver outer OMA in dBm,
$L_\mathrm{post}$ the downstream loss and $M$ the engineering margin in dB.
The factor $-10\log_{10}k_\mathrm{OMA}$ represents the electro-optic modulation
conversion penalty of the microring. Because reverse bias shifts the resonance
over only a fraction of its linewidth ($\Delta x \approx 0.23$), the optical transmission
swings between $T_0 \approx 0.38$ and $T_3 \approx 0.55$, converting approximately
$16.8\%$ of the input continuous-wave power into peak-to-peak modulated outer OMA.
This $7.7\text{--}8.0$~dB conversion penalty is distinct from the receiver's
$10\log_{10}3 \simeq 4.8$~dB PAM4 inner-eye slicing penalty, which arises from
subdividing the outer optical swing into three equal decision eyes at the slicer.

The IEEE 200G optical receiver baseline at 106.25~GBaud PAM4
(212.5~Gb/s gross) specifies $S_\mathrm{RX}=-3.2$~dBm for
low TECQ~\cite{welch_ieee2024}.
Table~\ref{tab:linkbudget} summarizes the optical link budget under co-packaged
edge coupling, with 1.0~dB input and output coupler losses. Downstream passive
loss totals $L_\mathrm{post}=3.5$~dB, comprising 1.5~dB on-chip routing, 1.0~dB
output edge coupling and 1.0~dB external fiber and connector loss.
At the nominal 11.3~mW bus power baseline ($+10.5$~dBm), the link delivers
$-0.7$~dBm outer OMA to the receiver for the 256G lateral ring and
$-1.0$~dBm for the Z-shaped ring, providing healthy unallocated implementation
margins of 2.5 and 2.2~dB, respectively.
With 1.0~dB input edge coupling, the continuous-wave laser output requirement
is 14.2~mW ($+11.5$~dBm) per lane, comfortably within commercial DFB laser ratings.

\begin{table}[t]
\centering
\caption{Per-lane optical link budget with 1.0~dB edge coupling at 11.3~mW bus power.}
\label{tab:linkbudget}
\footnotesize
\setlength{\tabcolsep}{6pt}
\renewcommand{\arraystretch}{0.98}
\begin{tabular}{lccc}
\toprule
\textbf{Link stage (laser $\to$ receiver)} & \textbf{Loss (dB)} & \textbf{256G Lateral (dBm)} & \textbf{Z-shaped (dBm)} \\
\midrule
Laser source output                        & ---         & $+11.5$ & $+11.5$ \\
Input edge coupler                         & $1.0$       & $+10.5$ & $+10.5$ \\
CW bus power ($P_\mathrm{bus} = 11.3$~mW)  & ---         & $\mathbf{+10.5}$ & $\mathbf{+10.5}$ \\
Modulator OMA conversion ($k_\mathrm{OMA}$)& $7.7$ / $8.0$ & $+2.8$  & $+2.5$  \\
On-chip routing                            & $1.5$       & $+1.3$  & $+1.0$  \\
Output edge coupler                        & $1.0$       & $+0.3$  & $0.0$   \\
External fiber channel                     & $1.0$       & $-0.7$  & $-1.0$  \\
Received outer OMA at TP3                  & ---         & $\mathbf{-0.7}$ & $\mathbf{-1.0}$ \\
Receiver sensitivity ($S_\mathrm{RX}$)     & ---         & $-3.2$  & $-3.2$  \\
\midrule
\textbf{Implementation margin ($M$)}       & ---         & $\mathbf{+2.5}$~dB & $\mathbf{+2.2}$~dB \\
\bottomrule
\end{tabular}
\end{table}

\subsection{Heating at the selected operating point}
At 11.3~mW bus input and held $x=-0.80$, the assumed 51\% CW
absorbed fraction gives 5.8~mW absorption and a 920~pm redshift.
The shift spans 1.8 linewidths, far exceeding the 59~pm drive excursion.
Under modulation, average absorption determines the thermal shift through
Eq.~(\ref{eq:opticalheat}), while the shifted resonance changes absorption
through Eq.~(\ref{eq:thermal_spectrum}). The hot state must therefore be
determined self-consistently.

\begin{figure}[t]
 \centering
 \includegraphics[width=\textwidth]{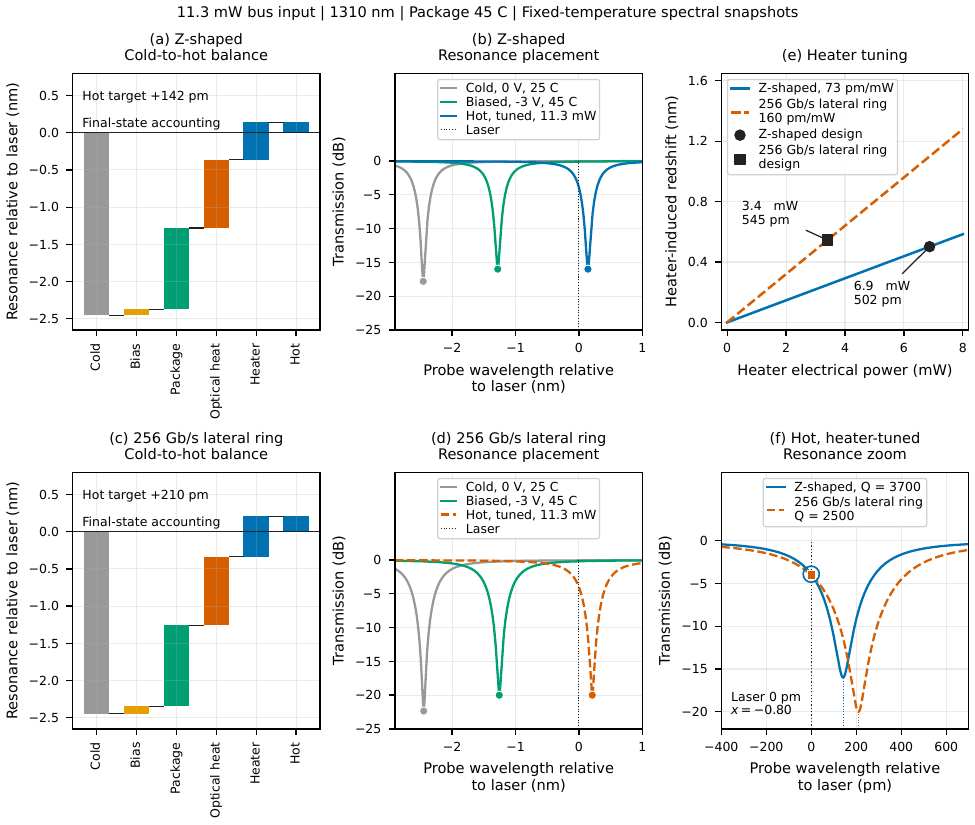}
 \caption{Conditional resonance placement at 11.3~mW, 1310~nm and
 45~$^\circ$C. \textbf{(a,c)}~Z-shaped ring/256G lateral ring wavelength balances.
 \textbf{(b,d)}~Corresponding frozen-temperature spectra.
 \textbf{(e)}~Heater-only shifts from calibrated efficiencies.
 \textbf{(f)}~Hot-spectrum zoom at the assumed $x=-0.80$ target.
 Circles/squares denote Z-shaped ring/256G lateral ring operating points; dotted lines
 mark the laser.}
 \label{fig:zthermal}
\end{figure}

The comparison of our Z-shaped ring and the 256G lateral ring in Fig.~\ref{fig:zthermal} uses
the same 11.3~mW nominal bus power for both rings. Both models assume
160~pm per milliwatt absorbed
and 54.6~pm per kelvin, with
$\overline{P}_\mathrm{abs}\simeq0.84P_\mathrm{bus}/(1+x^2)$.
Cold placement and heater power are jointly optimized for hot
$x=-0.80$ over 10.2--12.4~mW bus power, 40--50~$^\circ$C package temperature
and $\pm100$~pm fabrication error, with 0.5~mW reserve at each heater
limit. The assumed heater-power range for the 256G lateral ring is 0--8~mW. The
Z-shaped ring's heater-power range is 0--32~mW, corresponding to the demonstrated
0--4~V sweep at 500~$\Omega$~\cite{yuan_nc2024}.

At 1310~nm, the models of the Z-shaped ring and the 256G lateral ring use $Q_L=3700$ and 2500,
with 16~dB notch ER assigned to the Z-shaped ring's $-3$~V reference. Their
target resonances lie 142 and 210~pm red of the laser. The same 920~pm
self-shift spans 2.6 linewidths of the Z-shaped ring but 1.8 of the 256G lateral ring,
giving different feedback branches despite equal absorbed power.
Optimized cold resonances about 2.5~nm blue of the laser at 0~V and
25~$^\circ$C require minimum nominal heater powers of 6.9 and 3.4~mW,
respectively, with corner ranges of 0.5--13.2 and 0.5--6.3~mW.

For the 256G lateral ring, the hottest, highest-power and redmost fabrication
corner sets the lower heater reserve, limiting how far the cold resonance
can move red to reduce nominal heating. Laser turn-on with the heater
off gives only 70~pm self-shift. Heater tuning moves the resonance
toward the laser, increasing absorption to 5.8~mW and the self-shift
to 920~pm [Fig.~\ref{fig:zthermal}(c,d)]. Heater and optical heating
therefore act together during acquisition.

For the Z-shaped ring, applying the final 6.9~mW command directly selects
a stable branch 684~pm blue of the laser. A temporary 16~mW command
moves the resonance through the turning point. Returning to 6.9~mW
after that transition reaches the target 142~pm red of the laser.
The startup pulse thus selects the operating branch. The hot spectra in
Fig.~\ref{fig:zthermal} are fixed-temperature probe sweeps with
bias-dependent loss, with loss and coupling otherwise temperature
independent.

\subsection{Heater tuning}\label{sec:heater}
Joule heating raises the waveguide temperature and effective index,
shifting the resonance red. For constant heater tuning efficiency,
\begin{equation}
 \Delta\lambda_\mathrm{res}
     =\left.\frac{\Delta\lambda_\mathrm{res}}{P_h}\right|_\mathrm{ref}P_h.
 \label{eq:heatershift}
\end{equation}
where $P_h=V_h^2/R_h$ is heater dissipation, $V_h$ its voltage and
$R_h$ its resistance. The reference efficiency is fitted relative to
zero heater power. The heater efficiencies of the Z-shaped ring and the 256G lateral ring are
73 and 160~pm/mW, with resistances of 500 and 317~$\Omega$,
respectively~\cite{yuan_nc2024,xue_ofc2025}. At fixed loss and coupling,
heater tuning shifts the transmission curve in Eq.~(\ref{eq:xnorm})
without changing its shape.

\subsection{Cold resonance placement}\label{sec:coldplace}
Designed resonance spacing can reduce postfabrication tuning
requirements~\cite{krishnamoorthy_jphot2011}, while self-heating changes
the operating wavelength and branch acquisition~\cite{decea_oe2019}.
Cold placement must account for package temperature, DC bias and optical
heating while retaining heater correction in both directions. Ring path
length and cross-section set the cold wavelength, whereas coupling and
doping set the required hot-state linewidth, notch depth, tuning efficiency
and electrical response.

Relative to the 0~V, 25~$^\circ$C cold state without optical or heater
power, the first-order wavelength contributions add to give
\begin{equation}
 \begin{aligned}
    \Delta\lambda_\mathrm{res}
    ={}&\underbrace{\lambda_{\mathrm{res},V_\mathrm{DC}}-\lambda_{\mathrm{res},V_\mathrm{ref}}}_{\text{bias}}
          +\underbrace{\left.\Delta\lambda_\mathrm{res}\right|_{\Delta T}}_{\text{package}} 
         +\underbrace{\left.\Delta\lambda_\mathrm{res}\right|_{P_h}}_{\text{heater}}
          +\underbrace{\left.\Delta\lambda_\mathrm{res}\right|_{\overline{P}_\mathrm{abs}}}_{\text{self-heating}}
          +\underbrace{\left.\Delta\lambda_\mathrm{res}\right|_\mathrm{other}}_{\text{other}},
 \end{aligned}
 \label{eq:ladder}
\end{equation}
where $\lambda_\mathrm{res}-\lambda_\mathrm{laser}=c_0+\Delta\lambda_\mathrm{res}$,
$c_0$ is the cold offset from the laser, $V_\mathrm{ref}=0$ the
cold-reference junction voltage, $V_\mathrm{DC}$ the signed operating bias,
and $\Delta T$ the package-temperature rise above 25~$^\circ$C.
Each contribution is evaluated with the other sources fixed. Package,
heater and self-heating shifts follow Eqs.~(\ref{eq:thermoopticshift}),
(\ref{eq:heatershift}) and (\ref{eq:opticalheat}), respectively.
The bias-induced shift uses voltage-indexed wavelengths at the same cold
thermal reference. The residual $\left.\Delta\lambda_\mathrm{res}\right|_\mathrm{other}$
includes additional nonlinear-index and crosstalk shifts, such as those
from photocarriers, not already included in the named contributions.
The residual is set to zero in the numerical balance below.
Modulation-induced chirp follows from the time-dependent output phase of
the cavity model, so it need not be added as a separate resonance shift.
At the assumed 33~pm/V tuning efficiency, changing from 0 to $-3$~V gives a
$+99$~pm redshift. Package warming from 25 to 45~$^\circ$C adds $+1092$~pm, while
11.3~mW bus input contributes $+920$~pm of optical self-heating. Balancing these
shifts against a designed cold resonance of $-2447$~pm requires a nominal heater
shift of $+545$~pm (3.4~mW) to reach the target hot resonance at $+210$~pm
($x = -0.80$).
Separate heater and optical-heating states retain their response times
in the cavity model (Sec.~\ref{sec:va}).

In this cold design, self-heating reduces
the heater requirement from 9.2~mW without optical heating to 3.4~mW
at 11.3~mW bus input, assuming equal optical and heater tuning
efficiencies. The modulation excursion is evaluated about this settled
hot resonance. In physical mask layout, setting this cold resonance requires 
adjusting the drawn ring radius by
\begin{equation}
  \Delta R = \frac{L_\mathrm{rt}}{2\pi}\left(\frac{n_g}{n_\mathrm{eff}}\right)\left(\frac{c_0}{\lambda_\mathrm{res}}\right),
  \label{eq:radiuscoldshift}
\end{equation}
which evaluates to $\Delta R \approx -17$~nm for the 256G lateral ring and $-34$~nm for the Z-shaped ring. These sub-35~nm offsets fall comfortably within standard immersion-lithography mask grids, exploiting package warming and optical self-heating to eliminate over 12~mW of steady-state electrical heater dissipation per lane. At 256~Gb/s, this 12.6~mW reduction saves 49.1~fJ/bit, lowering active heater overhead from 62.4 to 13.3~fJ/bit and bringing thermal energy dissipation within the same order of magnitude as the electrical driver. For the 200~Gb/s Z-shaped ring, harnessing the same passive thermal assist saves 27.6~mW of equivalent heating, delivering a 137.8~fJ/bit energy-efficiency improvement.

Correction reserve sets the nominal heater command. For the 11.3~mW
example, $\pm5$~K package drift, $\pm100$~pm fabrication offset and
$\pm10\%$ bus-power variation give a maximum positive shift error of
465~pm. Compensating this error requires 2.9~mW downward adjustment,
which, with 0.5~mW reserve, sets the 3.4~mW nominal command.
The opposite corner requires 6.3~mW, within the 0--8~mW heater range
with 0.5~mW upper reserve. Cold placement therefore sets the steady
heater power and correction range, while the startup command determines
access to the intended branch.


\section{Compact modeling and eye-diagram evaluation}\label{sec:va}


\subsection{Physical and Verilog-A models}
Commercial spatial solvers characterize optical fields and carrier
transport, while circuit tools use compact component
models~\cite{interconnect_docs2026}. Compact models couple extracted
electrical parasitics to nonlinear cavity and thermal dynamics for
electronic--photonic co-design~\cite{bao_jlt2024}. For the segmented Z-shaped
and 256G lateral rings, measured or spatially simulated parameter relations
connect junction design, coupling and bias to OMA, bandwidth and thermal
power in the physical model.

For numerical integration in Python or Verilog-A, Eq.~(\ref{eq:cmt_mrm})
is expressed in the laser-frequency frame.
Writing $a(t)=\widetilde{a}(t)e^{j\omega_\mathrm{in}t}$ and
$E_{\mathrm{in,out}}(t)=\widetilde{E}_{\mathrm{in,out}}(t)e^{j\omega_\mathrm{in}t}$
for a constant reference frequency $\omega_\mathrm{in}$ gives
\begin{equation}
 \frac{d\widetilde{a}}{dt}
   =\left[j\Delta\omega-\frac{1}{2\tau_p}\right]\widetilde{a}
     +\frac{\widetilde{E}_\mathrm{in}}{\sqrt{\tau_e}},
 \qquad
 \widetilde{E}_\mathrm{out}
   =\widetilde{E}_\mathrm{in}-\frac{\widetilde{a}}{\sqrt{\tau_e}},
 \label{eq:cmt_envelope}
\end{equation}
where the tildes denote complex envelopes and
$\Delta\omega=\omega_r-\omega_\mathrm{in}$ is the signed detuning.
The envelopes preserve energy, port power, phase and cavity memory
without resolving the optical carrier.

\begin{figure*}[p]
  \centering
  \includegraphics[width=\textwidth,height=0.85\textheight,keepaspectratio]{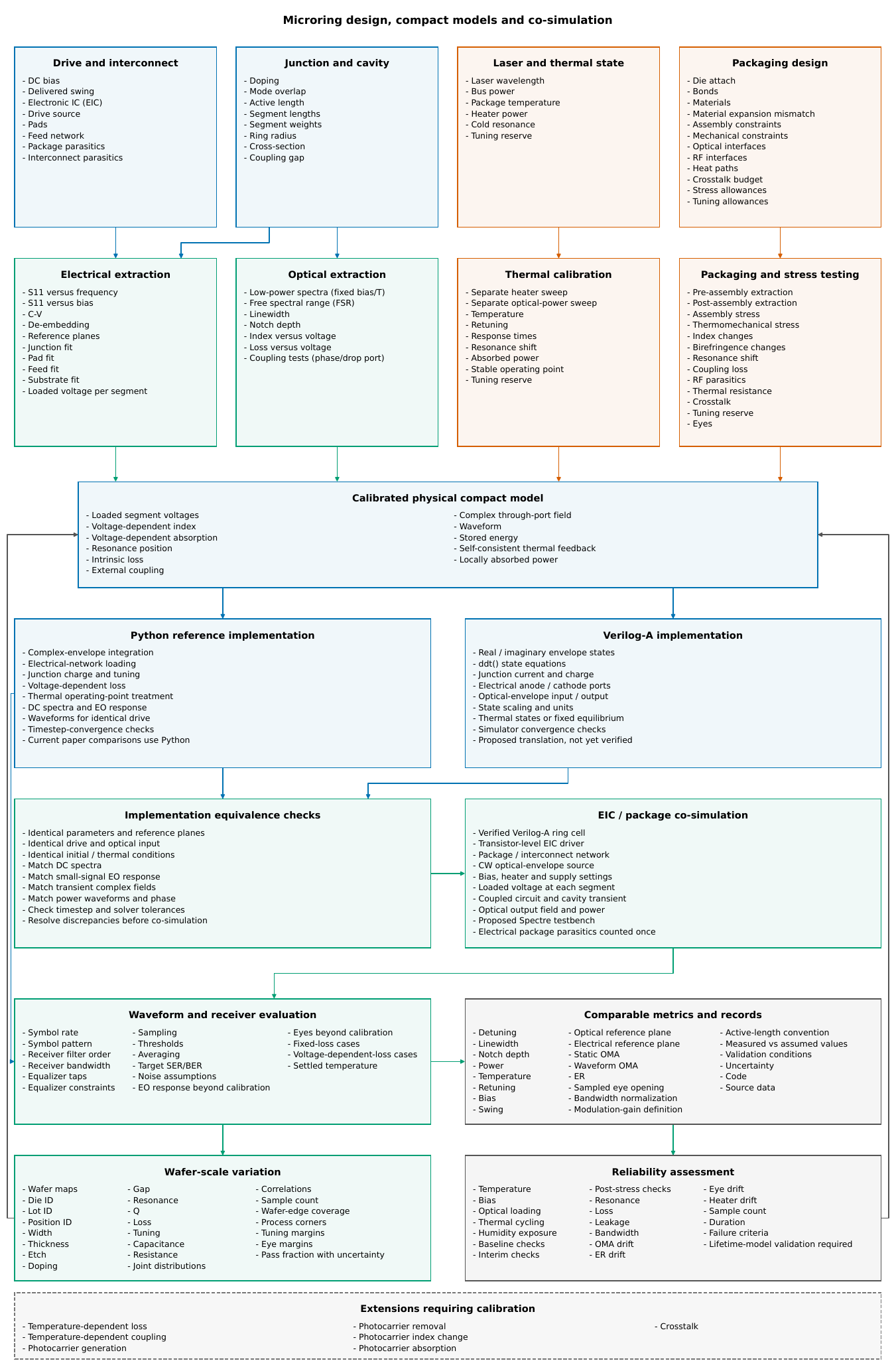}
  \caption{Microring design, compact models and co-simulation workflow.
  The dashed box denotes extensions requiring calibration.}
  \label{fig:mrmphysicaldesign}
\end{figure*}

Junction resistance, capacitance and the driving circuit determine the
voltage across each segment [Fig.~\ref{fig:mrmphysicaldesign}]. Depletion
changes the index and absorption, shifting the resonance and modifying
its linewidth and depth through Eqs.~(\ref{eq:loadeddecay}) and
(\ref{eq:notchdepth}). The loss-aware eyes retain voltage-dependent
intrinsic loss at fixed external coupling. Temperature-dependent
parameters and dynamic photocarrier effects require additional calibration.

The resonance combines instantaneous voltage tuning with the thermal
shifts in Eq.~(\ref{eq:ladder}). Separate first-order heater and
optical-heating states relax toward Eqs.~(\ref{eq:heatershift}) and
(\ref{eq:opticalheat}), with absorption determined by stored cavity
energy and locally absorbing loss. Thermal relaxation is much slower
than a symbol period, so the eye calculations use the equilibrium in
Eq.~(\ref{eq:thermal_spectrum}) at average absorption. The squared
magnitude of the through-port field gives optical power, including
interference, from which OMA, ER and sampled eye opening are extracted.

Verilog-A can express the real and imaginary envelope equations using
its built-in time-derivative operator \texttt{ddt()}. The circuit simulator
solves these differential equations together with the junction current
and charge relations and the transistor-level EIC and package network.
Matching spectra and waveforms
under identical conditions would verify the translation. The measurement
comparisons here use Python.

\subsection{Device calibration and design tolerance}
\label{sec:benchmarkconditions}
De-embedded complex S-parameters constrain the electrical network, as in
the Z-shaped ring model
[Fig.~\ref{fig:yuanmeasuredrf}] and bias-dependent circuit
extraction~\cite{yuan_nc2024,bao_jlt2024}. For real reference impedance
$Z_0$, the input impedance follows $Z_\mathrm{in}=Z_0(1+S_{11})/(1-S_{11})$.
Accurate circuit-level co-design requires de-embedding reference planes to the
junction boundary to decouple pad capacitance $C_\mathrm{pad}$ and feed
inductance $L_p$ from the active junction impedance $R_s$ and $C_j(V)$.
Un-deembedded one-port measurements lump packaging and pad shunts into the
junction, masking high-frequency inductive peaking and distorting the extracted
capacitance. Multi-length test structures, such as standalone test diodes and
ring segments with varying active arc lengths, resolve correlated parameters
between junction sheet resistance and metal contact resistance, recovering the
true junction-voltage transfer function $H_e(\omega_m)$ in Eq.~(\ref{eq:vjvsrc}).

Complementary optical parameter extraction must be performed in the linear,
weak-probe regime (input power below $-10$~dBm) to prevent optical self-heating
from distorting the extracted linewidth and notch depth. Group index $n_g$,
loaded quality factor $Q_L$, and notch depth are determined from through-port
transmission via Eqs.~(\ref{eq:fsr}), (\ref{eq:static}) and (\ref{eq:notchdepth}).
Because through-port intensity transmission exhibits identical Lorentzian dips for
undercoupled and overcoupled cavities, designers must resolve the coupling regime
unambiguously by measuring transmission phase, probing an auxiliary drop port, or
tracking the notch depth derivative under reverse bias. Voltage sweeps then
constrain the electro-optic phase efficiency $\eta(V)$ and voltage-dependent
absorption $\Delta\alpha(V)$, while thermal steps constrain heater tuning
efficiencies and thermal response times for large-signal Verilog-A tables.

Translating device-level calibration to multi-lane DWDM arrays requires managing
dimensional fabrication tolerances. In silicon photonics foundry processes,
across-wafer waveguide width and thickness exhibit typical standard deviations
of $\sigma_w \approx 4.6$~nm and $\sigma_h \approx 0.8$~nm, with systematic wafer-scale
gradients accounting for roughly 80\% of the total variance~\cite{xing_acsphotonics2023}.
Because each nanometer of width deviation shifts the resonance by approximately
1~nm, uncompensated fabrication skew can exceed the entire channel spacing of a
dense DWDM grid. Array layouts must therefore place microring modulators in
immediate spatial proximity with matched dummy pattern fills and identical orientation.
This tight layout matching suppresses intra-array relative skew to below 100~pm,
confining active thermal tuning to common-mode tracking rather than wide per-channel
correction.

Passive tolerance engineering can substantially desensitize the cavity to etching
and lithographic variations before active compensation. Incorporating adiabatically
widened waveguide bends in non-modulating arcs reduces optical mode interaction with
etched sidewall roughness, cutting across-wafer resonance wavelength standard deviation
by a factor of 2.1 relative to uniform narrow rings~\cite{mikkelsen_ring_tolerance2014}.
In the bus coupling region, shallow-etched ridge couplers provide up to a fourfold
reduction in the normalized standard deviation of the coupling coefficient compared to
deep-etched strip couplers~\cite{mikkelsen_coupler_tolerance2014}, stabilizing the power
transfer fraction $\kappa^2$ against over- or undercoupling drift. For active modulators,
width transitions between widened passive arcs and narrow active junctions must remain
strictly adiabatic to preserve fundamental mode purity and avoid modal scattering loss.

Statistical yield must be evaluated at the link level rather than through isolated
component tolerances~\cite{robinson_oe2025_variation}. In dense co-packaged transmitters,
the available on-chip heater power (typically 0 to 8~mW per lane) sets a strict upper
bound on correctable resonance error. Monte Carlo co-simulation should jointly sample
the correlated distributions of junction capacitance, series resistance, cold resonance
offset and tuning efficiency against receiver sensitivity and link loss. To prevent thermal
runaway or excessive interchannel thermal crosstalk, transceivers should implement
algorithmic wavelength arbitration, such as cyclic locking or nearest-line assignment,
which remaps logical data streams to adjacent optical carrier lines and cuts worst-case
thermal tuning power by more than 60\% compared to fixed-grid locking.

Design-for-test architectures must separate fabrication variations from
packaging-induced assembly shifts. Dedicated grating couplers enable automated
wafer-level optical probing to screen cold resonance wavelengths, loaded quality
factors and DC extinction ratios before committing dies to 2.5D or 3D integration~\cite{tan_foe2023}.
Subsequent assembly introduces parasitic shifts, including 20 to 50~pH of micro-bump
or wirebond inductance and thermal dissipation from adjacent CMOS driver dies.
Characterizing test structures before and after packaging isolates package-induced
electrical loading from intrinsic silicon process drift, confirming that the high-speed
eye margin remains intact.

During live operation, closed-loop feedback stabilizes the modulator against
ambient temperature drift~\cite{padmaraju_oe2012}. Control loops monitoring through-port
average power or dedicated tap photodiodes dynamically trim micro-heaters to lock
the operating detuning near the target optimum ($x \approx -0.80$). The feedback algorithm
must account for the asymmetric thermal stability between the self-stabilizing blue-detuned
flank and the regenerative red-detuned flank to prevent lock runaway. Over operating
lifetimes, high intra-cavity optical intensity and sustained reverse-bias electric
fields can induce charge trapping in the oxide or slab interfaces, causing slow threshold
shifts. Transceiver co-design should allocate a 10\% to 15\% overdrive margin in the
CMOS driver swing to absorb long-term efficiency degradation without violating optical
modulation amplitude specifications.

\subsection{Validation and eye-quality analysis}
\subsubsection{Measured and modeled responses}
\label{sec:yuantransmission}
\begin{figure*}[t]
 \centering
 \includegraphics[width=\textwidth]{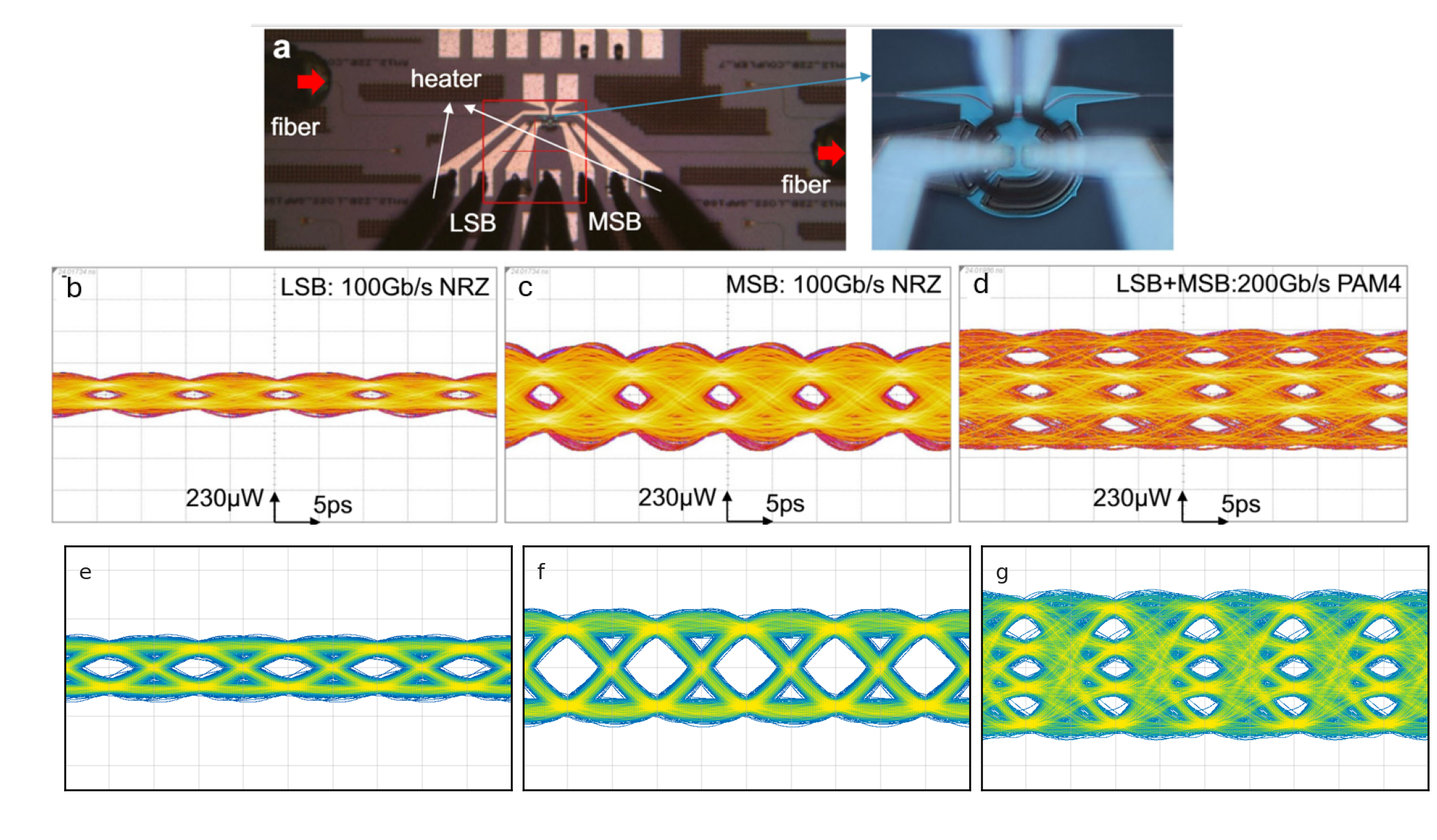}
  \caption{Z-shaped ring transmission. \textbf{(a)}~Device illustration and micrograph.
 \textbf{(b,c)}~Measured 100~Gb/s LSB/MSB NRZ eyes. \textbf{(d)}~Measured
 200~Gb/s PAM4. \textbf{(e--g)}~Simulated LSB, MSB and PAM4 eyes with noise
 and 21-tap FFE, without measurement averaging. Panels (a)--(d) cropped/rescaled from
 Yuan \emph{et al.}, Fig.~4~\cite{yuan_nc2024},
 \href{https://creativecommons.org/licenses/by/4.0/}{CC BY 4.0}.}
 \label{fig:yuanmeasuredeyes}
\end{figure*}

For our Z-shaped ring, the model with its electrical network extracted from segment
$S_{11}$ [Fig.~\ref{fig:yuanmeasuredrf}] gives 49~GHz EO bandwidth, matching the
measured value~\cite{yuan_nc2024}. The 200~Gb/s eyes in
Fig.~\ref{fig:yuanmeasuredeyes} compare measured and modeled responses.

The measured PAM4 eyes use two 100~Gb/s NRZ drives at
1.6~V$_\mathrm{pp}$ per segment after the bias-tees. The measured outer ER
is 3.6~dB and TDECQ is 0.2~dB at an SER threshold of $10^{-2}$ with a
21-tap FFE, applying Bessel filtering and 64-fold pattern
averaging to suppress amplified spontaneous emission (ASE)~\cite{yuan_nc2024}.

The simulations use equal 1.6~V$_\mathrm{pp}$ drives and an assumed
9/18~pm/V tuning split motivated by approximately two-to-one geometry
and capacitance.
Panels (e--g) use single acquisitions with noise and 21-tap FFE, without
pattern averaging. One amplitude calibration to the measured LSB rail
separation is shared across simulations at 230~$\mu$W and 5~ps per
division. Differences in processing limit comparisons of absolute rail spacing.

\subsubsection{Detuning and eye quality}
\begin{figure*}[t]
 \centering
 \includegraphics[width=\textwidth]{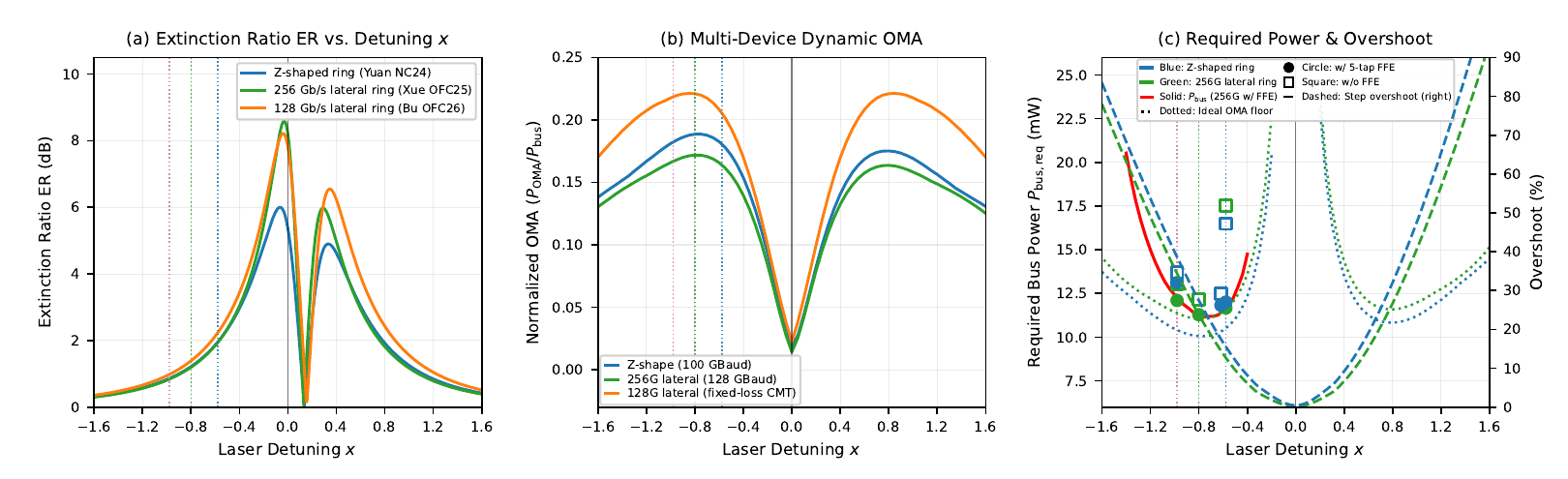}
 \caption{Detuning trade-offs across extinction ratio, optical modulation amplitude, and link power. \textbf{(a)}~Static two-rail extinction ratio ER. \textbf{(b)}~Large-signal dynamic OMA across three reference devices. \textbf{(c)}~Required bus power $P_\mathrm{bus,req}$ and limiting step overshoot under voltage-dependent loss. Dotted vertical lines mark maximum slope ($x = -0.58$, blue), maximum dynamic OMA ($x = -0.80$, green), and circuit-loaded maximum GBW ($x = -0.98$, pink).}
 \label{fig:bandwidthovershoot}
\end{figure*}

Increasing detuning broadens the optical response at the cost of low-frequency gain
and greater overshoot. The small-signal response is shown in
Fig.~\ref{fig:eoresponse}, where cavity peaking grows with detuning while
the loaded electrical response of Eq.~(\ref{eq:vjvsrc}) limits the cascade, so
the modeled electro-optic bandwidth shows no net peaking.
Figure~\ref{fig:bandwidthovershoot} connects this detuning dependence to large-signal
metrics and time-domain distortion across laser detuning $x$.
Static extinction ratio [Fig.~\ref{fig:bandwidthovershoot}(a)] increases as the
laser approaches the resonance notch. Because reverse bias widens depletion and
red-shifts the resonance ($\Delta x > 0$), it drives the lower rail directly into
the notch on the red side ($x > 0$), producing a steeper static extinction ratio
than on the blue side.
Dynamic optical modulation amplitude under voltage-dependent loss
[Fig.~\ref{fig:bandwidthovershoot}(b)] instead favors the blue side ($x < 0$),
peaking near $x \approx -0.58$ for the Z-shaped ring and $x \approx -0.80$ for the
256G lateral ring. Here, resonance red-shift and reduced free-carrier
absorption combine constructively to increase transmission, whereas on the red side
competing effects reduce dynamic OMA by 5\% to 8\%.
Figure~\ref{fig:bandwidthovershoot}(c) carries this trade-off into the link domain,
comparing required optical bus power $P_\mathrm{bus,req}$ on the left axis against limiting
step overshoot on the right axis. Junction parasitics shift these operating points relative
to the bare optical cavity. At DC, vanishing capacitive displacement current leaves the
maximum-slope point locked at the stationary Lorentzian inflection point ($x = -0.58$).
Under 128~GBaud modulation, the dynamic OMA peak shifts outward to $x \approx -0.80$ as
optical cavity peaking compensates for junction $RC$ roll-off, while electrical
filtering pulls the maximum-GBW bias inward from the bare-cavity value ($|x| = 1.13$) to $x = -0.98$.

Step overshoot reaches 24\% for the 256G lateral ring and 27\% for the Z-shaped ring at $x = -0.80$,
increasing to 36\% at $x = -0.98$. Required bus power forms a bathtub minimum near
$x \approx -0.80$, reaching 11.3~mW for the 256G lateral ring and 11.8~mW for the
Z-shaped ring. Dotted curves trace the ideal-OMA floor without intersymbol interference.
Without receiver equalization, intersymbol interference increases the power required at
maximum slope to 17.5~mW (squares), whereas a five-tap equalizer lowers required power to
11.7~mW (circles), providing a 5.8~mW equalization recovery. 

The detuning trade-off also affects level spacing and noise tolerance.
The noise comparisons use the fixed-loss 256G lateral-ring model
(Sec.~\ref{sec:ringdyn}), with a refined GBW bias at
$x=-0.98$. Table~\ref{tab:eyequality} gives the sweep at 11.3~mW bus input.
Peaking and rise metrics use small-signal
and step responses, while OMA and ER use waveform extrema. Despite greater
overshoot at the bare-cavity GBW bias, its five-tap equalized penalty is
close to that at maximum slope. Fitted level linearities are 97\% at the
bare-cavity GBW bias and 89\% at maximum slope. Rise times are percentages of the unit interval
(UI), the symbol duration.

\begin{table}[t]
\centering
\caption{Fixed-loss eye properties at 128~GBaud and 1.8~V$_\mathrm{pp}$, normalized to 11.3~mW bus input. Penalties are not compliance TDECQ.}
\label{tab:eyequality}
\small
\setlength{\tabcolsep}{4.5pt}
\renewcommand{\arraystretch}{1.2}
\begin{tabular}{lccccccc}
\toprule
\textbf{Operating point} & $|x|$ & \textbf{Peaking} & \textbf{Overshoot} & \textbf{Rise}
 & \textbf{ER} & \textbf{OMA} & \textbf{TDECQ} \\
 & & (dB) & (\%) & (\% UI) & (dB) & (\% $P_\mathrm{in}$) & ($10^{-2}$ dB) \\
\midrule
Maximum small-signal slope & 0.58 & 0.3 & 13 & 38 & 2.8 & 16 & 8 \\
Reported bias                & 0.98 & 2.5 & 36 & 19 & 1.5 & 16 & 11 \\
Near bare-cavity GBW optimum & 1.13 & 3.5 & 47 & 16 & 1.2 & 16 & 14 \\
Beyond the optimum           & 1.50 & 6.2 & 76 & 9 & 0.9 & 14 & 18 \\
\bottomrule
\end{tabular}
\end{table}

Across the held-detuning sweep, waveform OMA varies by 7\% at 1.8~V$_\mathrm{pp}$
and 5\% at 4~V$_\mathrm{pp}$. Detuning directly affects link power
margin and level spacing, requiring thermal stabilization to preserve
the target operating point. The fixed-loss sweeps exclude capture dynamics,
power-dependent loss and carrier distortion.

\subsubsection{Noise loading and TDECQ}
TDECQ expresses the eye distortions above as a noise-tolerance penalty,
comparing ideal and device noise tolerances at the same outer OMA.
The simplified estimator is $10\log_{10}(\sigma_\mathrm{ideal}/\sigma_\mathrm{eq})$.
Here
$\sigma_\mathrm{ideal}=\mathrm{OMA_{outer}}/(6Q_t)$ and $\sigma_\mathrm{eq}$ is
the Gaussian noise RMS at the target SER. A fourth-order half-baud
Bessel--Thomson receiver precedes delay compensation and equalization,
with equalized level means setting thresholds and outer OMA. This
receiver and its five-tap symbol-spaced equalizer follow the
reference-receiver and reference-equalizer convention of the IEEE~802.3
transmitter test~\cite{ieee8023dj}. The fixed reference receiver isolates
transmitter quality and leaves the penalty independent of the photodiode
a link later uses. Compliance
TDECQ instead requires prescribed outer-level extraction, receiver,
equalizer constraints and impairment treatment.

For the fixed-loss 256G lateral ring, $Q_t=3.4$ and target SER
$4.8\times10^{-4}$ give five-tap penalties of 0.1~dB at
1.8~V$_\mathrm{pp}$ and 0.6~dB at 4~V$_\mathrm{pp}$.
Twenty-one taps change both by 0.01~dB.

Reported TDECQ values of 1.6 and 3.9~dB at 224 and 240~Gb/s
PAM4~\cite{sakib_ofc2022} were taken at different symbol rates of 112 and
120~GBaud, so the higher baud rate accounts for most of the gap, with
receiver filtering a secondary effect. Because TDECQ depends on the reference
receiver and equalizer prescribed by each interface standard, the two numbers
are compliance results under different specifications and are not directly
comparable.

Receiver filtering can change the reported penalty for an unchanged
optical waveform. At 128~GBaud, 1.8~V$_\mathrm{pp}$ and the reference bias
$x=-0.98$, the fixed-loss 256G lateral ring gives simplified penalties of 1.7 and
0.1~dB with fourth-order Bessel receivers at approximately 27 and 64~GHz.
Both use five unconstrained symbol-spaced least-squares taps and the same
phase and threshold algorithms, refitted after filtering with the same
record for fitting and evaluation. Increasing resolution from 32 to 64
samples per symbol changes either penalty by less than 0.1~dB. An identical-waveform comparison with
approximately 40 and 96~GHz receivers gives about 0.2~dB in both cases,
illustrating the dependence on the waveform and evaluation procedure.

\begin{figure*}[t]
  \centering
  \includegraphics[width=\textwidth]{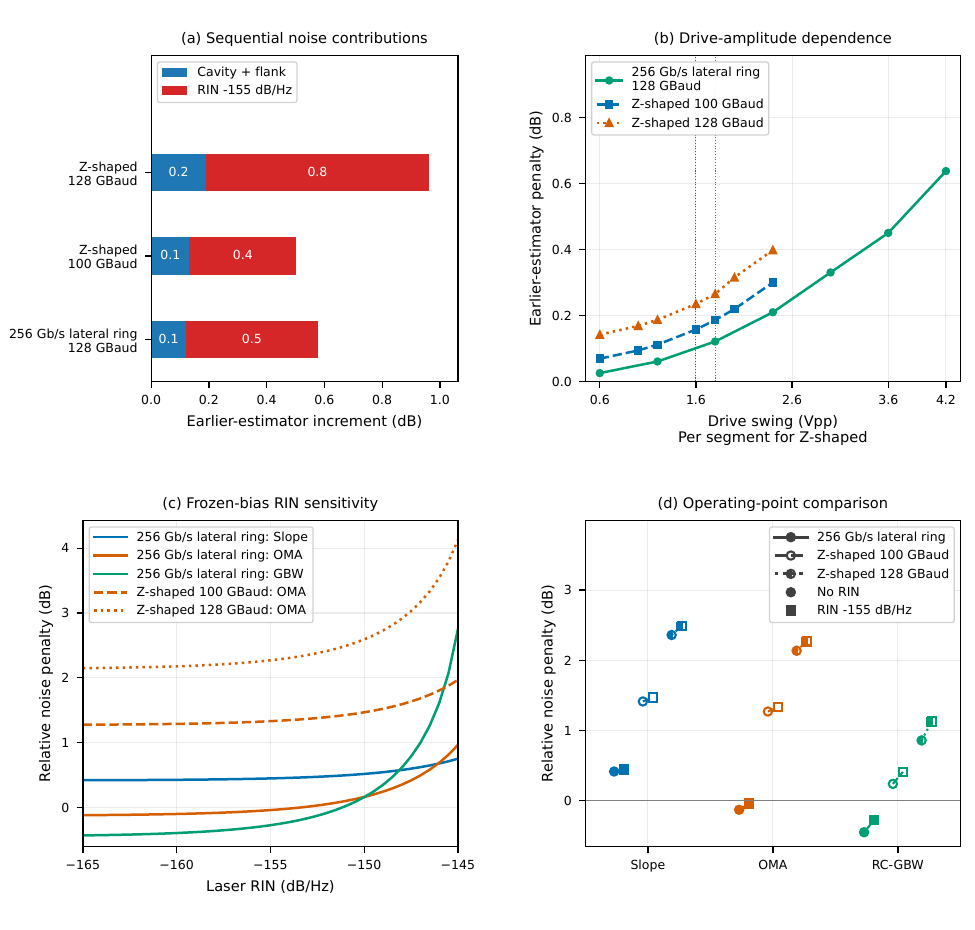}
  \caption{Modeled noise penalties, not compliance TDECQ.
  \textbf{(a,b)}~Cavity/RIN contributions and drive sweep at $|x|=0.98$,
  near the maximum GBW bias, for all three device/rate cases.
  \textbf{(c)}~RIN sensitivity at three 256G lateral-ring biases and maximum
  OMA for the Z-shaped ring at 100/128~GBaud.
  \textbf{(d)}~All three biases for each case, without and with
  $-155$~dB/Hz RIN. Panels (c,d) subtract each case's ideal reference;
  the Z-shaped ring results at 128~GBaud are model extrapolations.}
  \label{fig:tdecq}
\end{figure*}

Figure~\ref{fig:tdecq} compares the 256G lateral-ring model at
128~GBaud and 1.8~V$_\mathrm{pp}$ with the Z-shaped ring model at 100 and
128~GBaud and 1.6~V$_\mathrm{pp}$ per segment. The latter uses $Q_L=3700$, fitted
segment circuits with a 50~$\Omega$ source, an assumed 9/18~pm/V tuning
split and PRBS9 streams, versus random PAM4 for the 256G lateral ring.
Both use fixed intrinsic loss, half-baud receivers and five-tap equalizers.
Panels (a,b) use the estimator above at the common reference bias $x=-0.98$
for all three device/rate cases. Panel (a) adds multiplicative output RIN, while the drive sweep in
panel (b) excludes RIN. Panels (c,d)
propagate input RIN and subtract the ideal-reference penalty for each
device and rate. Panel (c) retains all three 256G lateral-ring biases and only
the maximum-OMA bias for the Z-shaped ring at each rate. Panel (d) includes all three
biases for each case. The drive sweep extrapolates the fixed-loss model.

Sequentially adding the electrical, cavity and RIN responses to
rectangular PAM4 quantifies their incremental noise penalties
[Fig.~\ref{fig:tdecq}(a)]. At 128~GBaud, the 256G lateral ring's internal
electrical filtering contributes less than 0.01~dB,
versus 0.1~dB from the coupled cavity memory and flank curvature and
0.5~dB from output RIN. For the Z-shaped ring at 100~GBaud, the corresponding cavity
and RIN increments are 0.1 and 0.4~dB. Their 0.5~dB sum excludes the
0.024~dB electrical-only increment. At 128~GBaud,
the Z-shaped ring's cavity and output-RIN increments increase to 0.2 and 0.8~dB,
respectively, again excluding the electrical-only increment. RIN dominates
the cavity contribution under these assumptions.

At the nominal drives above, moving from $x=-0.98$ to the maximum-dynamic-OMA
bias gives a larger optical modulation amplitude with a smaller output-RIN
penalty. The 256G lateral ring at 128~GBaud and the Z-shaped ring at 100 and
128~GBaud have maximum-OMA biases of $x=-0.80$, $-0.62$ and $-0.63$,
respectively. Their waveform-extrema OMA increases by 2.9\%, 12.0\% and
14.4\%, while the penalties without RIN decrease by
0.042, 0.015 and 0.011~dB with
the same estimator. At $-155$~dB/Hz output RIN, the summed cavity and RIN
penalties decrease from 0.6 to 0.3~dB, 0.5 to 0.2~dB and 1.0 to 0.5~dB,
respectively, retaining the electrical-only subtraction of panel (a).
The lower mean transmitted power at maximum OMA reduces multiplicative
output noise at fixed bus input power.

The Z-shaped ring has higher equalized PAM4 level uniformity at
$x=-0.98$. This uniformity, the level-separation-mismatch ratio
$R_\mathrm{LM}$ fitted from the four sampled power-level means after
receiver filtering and equalization, changes from 96.4\% to 92.5\% at
100~GBaud and from 97.8\% to 92.2\% at 128~GBaud when moved to maximum OMA. A broader
response can reduce intersymbol interference, while detuning also changes
the voltage-to-power curvature sampled by the four levels. The fitted
level spacing reflects both effects together with receiver filtering and
equalization. The common reference bias therefore provides higher equalized
PAM4 level uniformity, while maximum OMA provides greater optical modulation
amplitude and a lower output-RIN penalty under these conditions.

At the 256G lateral ring's fixed drive-sweep bias, optical bandwidth exceeds
twice Nyquist, leaving little residual ISI for additional taps to remove.
Increasing swing instead samples more Lorentzian curvature and compresses
outer levels. The equalized PAM4 level uniformity $R_\mathrm{LM}$ falls from
99\% to 96\%, approaching the 95\% comparison
criterion. The smaller drive reduces the penalty but gives less than half
the OMA, limiting tolerable link loss. Drive and bias must therefore meet
OMA, level-spacing and noise requirements together.

\subsubsection{Relative intensity noise}
Unlike the output-RIN loading above, the input-RIN comparison includes
cavity filtering. Figure~\ref{fig:tdecq}(c,d) compares three fixed-loss
operating points of the 256G lateral ring with a 64~GHz receiver and
five-tap symbol-spaced equalizer. Taps and thresholds fitted on the training
half are frozen during evaluation, with equalizer noise enhancement
included. Subtracting the 0.4~dB
filtered-ideal-PAM4 penalty gives signed values, with negative values
indicating improvement over that reference under unconstrained equalization.

White input RIN is propagated through the frozen-bias cavity, receiver
and equalizer over 0--64~GHz. At $-155$~dB/Hz, the signed penalties change from
0.42 to 0.45~dB at maximum slope,
from $-0.13$ to $-0.041$~dB at
maximum OMA, and from $-0.45$ to
$-0.28$~dB at maximum GBW
[Fig.~\ref{fig:tdecq}(d)]. GBW's larger increment is consistent with higher
mean transmission and lower contrast. The calculation excludes noise
above 64~GHz, aliasing and level-dependent conversion during modulation.

The Z-shaped ring comparison at 100~GBaud uses the undercoupled branch and both fitted
segment networks with a 50~$\Omega$ source, the same training/evaluation
split, a 50~GHz Bessel receiver and five-tap equalizer. Its own 0.4~dB
ideal-reference penalty is subtracted. At slope, dynamic-OMA and
circuit-loaded GBW biases of $|x|=0.58$, 0.62 and 0.97, the baseline
penalties are 1.4, 1.3 and 0.2~dB. Input RIN at $-155$~dB/Hz adds
0.056, 0.061 and
0.17~dB, respectively, over 0--50~GHz under the
frozen-bias approximation. The normalized comparison favors circuit-loaded
GBW at the demonstrated 100~GBaud rate, while link-power selection also requires absolute
OMA and sampled level separation.

At 128~GBaud, the same Z-shaped ring model uses a 64~GHz receiver with input
RIN integrated over 0--64~GHz. The slope, rate-specific waveform-OMA and
circuit-loaded GBW biases are $|x|=0.58$, 0.63 and 0.97, with relative
penalties of 2.4, 2.1 and 0.9~dB without RIN, rising to 2.5, 2.3 and
1.1~dB at $-155$~dB/Hz. The five-tap training/evaluation procedure is
unchanged. Increasing resolution from 64 to 128 samples per symbol changes
these penalties by less than 0.02~dB. The higher-rate case is a fixed-loss
model extrapolation, not the demonstrated 200~Gb/s (100~GBaud) Z-shaped ring experiment.

Absolute noise tolerance also depends on level separation. For the
256G lateral ring under the common calibration in
Sec.~\ref{sec:ratebudget}, maximum OMA requires the least power despite
GBW's lower normalized penalty, as shown by the required-power comparison
at 128~GBaud in that section.
Minimum equalized eyes are 90\%, 86\% and 88\% of the ideal adjacent-level
gap at OMA, slope and GBW, respectively. Absolute level separation,
residual ISI and RIN jointly determine sensitivity through
Eq.~(\ref{eq:rate_required_power}).

Transmitter pre-emphasis is also constrained by the available swing.
For the 256G lateral ring, peak-limited two-tap TX searches at
128~GBaud select bypass for OMA and GBW with five-tap RX FFE. At the
1.8~V$_\mathrm{pp}$ command limit, precursor/postcursor emphasis reduces
sustained drive during long symbol runs. Figure~\ref{fig:operatingrate}
retains bypass, with slope and higher-rate TX settings unoptimized.

\section{Conclusion and outlook}\label{sec:conclusion}

Interconnect scaling is leading the performance and energy footprint of distributed artificial intelligence fabrics. Depletion-mode silicon microrings have proven their capability by delivering 200~Gb/s PAM4 with record-low switching energy, yet scaling to 400G per lane and multi-Terabit escape densities demands a fundamental departure from isolated component design toward integrated electronic--photonic co-design.
Optical modulators and electronic transceivers must therefore be designed together as a single system. The physical and Verilog-A models developed in this review provide the unified framework to enable this co-design.

Scaling individual lanes toward 400G at 212.5~GBaud PAM4 requires balancing photon lifetime against resonant enhancement through controlled overcoupling and detuning-peaking. High-bandwidth avalanche photodiodes provide a complementary route to overcome electronic receiver noise under high-speed operating photocurrents. Across multi-channel transmitter arrays, dense wavelength-division multiplexing must scale concurrently through automated thermal locking and channel allocation to suppress interchannel crosstalk as aggregate bandwidth density increases.

Beyond device and wavelength scaling, high-speed optical links face an emerging gap between interoperability standards and real-world deployment needs. Specifications such as IEEE 802.3dj define baseline compliance thresholds, yet hyperscale datacenters often require pre-FEC bit-error rates near $10^{-9}$ to maintain mission-critical application performance. Quantifying and protecting this unallocated implementation margin represents an essential co-design problem for the industry.

Managing that margin over dynamic operating conditions points toward adaptive, closed-loop optical interfaces. Frameworks such as autonomous path startup (APSU) and inter-sublayer link training (ILT) can be extended from the electrical host interface into the optical domain. Using far-end receiver telemetry to dynamically adjust transmitter operating points creates an actionable route toward continuous link-margin tracking and in-service retraining without interrupting live data traffic.

The predictive modeling framework presented here bridges device physics with circuit and link design. By unifying physical device equations, Verilog-A circuit co-simulation, and link-budget evaluation, this framework equips Tier-1 network system architects to balance power, thermal, and reach trade-offs across distributed AI fabrics, while enabling photonic and electronic designers to systematically explore the multidimensional co-design space to scale optical interconnects to 400G per lane and multi-Terabit capacities.

\begin{acknowledgments}
The authors thank Ankur Kumar and Ruida Liu for EIC support, and Luca Ramini,
Charin Hong, and Wayne V. Sorin for helpful discussions on interconnect simulation and testing.
\end{acknowledgments}

\section*{Author Declarations}

\subsection*{Conflict of Interest}
The authors have no conflicts to disclose.

\end{document}